\documentclass[aps,prl,reprint,groupedaddress]{revtex4-1} 
\usepackage{graphicx}
\usepackage{siunitx}
\usepackage{amsmath,amssymb}
\usepackage{textcomp}
\usepackage{color}

\newcommand{\ket}[1]{\left\lvert #1 \right\rangle}%
\newcommand{\avg}[1]{\left\langle #1 \right\rangle}%
\newcommand{\abs}[1]{{\lvert #1 \rvert}}%

\begin{document}

\title{Quantum gas microscopy of an attractive Fermi-Hubbard system}

\author{Debayan Mitra}
\author{Peter T. Brown}
\author{Elmer Guardado-Sanchez}
\author{Stanimir S. Kondov}
\author{Trithep Devakul}
\author{David A. Huse}
\author{Peter Schau{\ss}}
\author{Waseem S. Bakr}
\email[]{wbakr@princeton.edu}

\affiliation{Department of Physics, Princeton University, Princeton, New Jersey 08544, USA}

\date{\today}

\begin{abstract}
The attractive Fermi-Hubbard model is the simplest theoretical model for studying pairing and superconductivity of fermions on a lattice \cite{Micnas1990}.  Although its $s$-wave pairing symmetry excludes it as a microscopic model for high-temperature superconductivity, it exhibits much of the relevant phenomenology, including a short-coherence length at intermediate coupling and a pseudogap regime with anomalous properties \cite{Randeria1992,Trivedi1995,Singer1996,Kyung2001}. Here we study an experimental realization of this model using a two-dimensional (2D) atomic Fermi gas in an optical lattice. Our site-resolved measurements on the normal state reveal checkerboard charge-density-wave correlations close to half-filling. A ``hidden" SU(2) pseudo-spin symmetry of the Hubbard model at half-filling guarantees superfluid correlations in our system \cite{Yang1990}, the first evidence for such correlations in a single-band Hubbard system of ultracold fermions. Compared to the paired atom fraction, we find the charge-density-wave correlations to be a much more sensitive thermometer, useful for optimizing cooling into superfluid phases in future experiments. \end{abstract}


\maketitle


The Fermi-Hubbard model is a fundamental condensed matter model for studying strongly-correlated fermions on a lattice \cite{Hubbard1963,Auerbach1990}. The numerical intractability of the model at low temperatures has spurred much experimental work on quantum simulations using repulsively interacting cold atoms in optical lattices \cite{Joerdens2008,Schneider2008,Greif2013,Hart2015,Parsons2016,Cheuk2016,Boll2016,Brown2016,Cocchi2016}.  In contrast, the attractive model has received much less attention from the cold atom community \cite{Strohmaier2007,Hackermueller2010,Schneider2012}, despite being an ideal playground for exploring short coherence-length superconductivity on a lattice. In the continuum, strongly-interacting attractive fermions have been studied using Feshbach resonances, resulting in the observation of superfluid gases across the crossover from molecular Bose-Einstein condensates (BEC) to Bardeen-Cooper-Schrieffer superfluids (BCS) \cite{Inguscio2008}. Fermionic superfluids have also been prepared in optical lattices close to a Feshbach resonance \cite{Chin2006}. However, these systems are not described by a simple Hubbard model due to multi-band couplings and off-site interactions \cite{Duan2005,Carr2005,Zhou2005,Diener2006}.


In this work, we focus on the 2D attractive Fermi-Hubbard model which has been theoretically studied in detail \cite{Hirsch1985,Scalettar1989,Micnas1990,Moreo1991,Paiva2004}. Our experiments are performed at an interaction energy small compared to the bandgap, where the single-band Hubbard description is applicable. In a grand-canonical ensemble, the Hamiltonian of the system is given by ${\cal {H}}=-t\sum_{\langle \textbf{rr}^\prime\rangle,\sigma} \left(c^{\dag}_{\textbf{r},\sigma}c_{\textbf{r}^{\prime},\sigma}+c^{\dag}_{\textbf{r}^{\prime},\sigma}c_{\textbf{r},\sigma}\right) + U \sum_\textbf{r} n_{\textbf{r},\uparrow}n_{\textbf{r},\downarrow}-\mu\sum_{\textbf{r},\sigma} n_{\textbf{r},\sigma}$. Here $c^{\dag}_{\textbf{r},\sigma}$ is the creation operator for a fermion with spin $\sigma$ on site $\textbf{r}$, $n_{\textbf{r},\sigma} = c^{\dag}_{\textbf{r},\sigma}c_{\textbf{r}\sigma}$, $t$ is the tunneling matrix element between nearest-neighbor lattice sites, $U < 0$ is the strength of the on-site interaction and $\mu$ is a spin-independent chemical potential. At low temperatures, the fermions undergo a Berezinskii-Kosterlitz-Thouless (BKT) transition to a superfluid phase. As $U/t$ is increased, the superfluid crosses over from a BCS-type superfluid to BEC of hardcore bosons, with the critical temperature reaching a maximum in the intermediate coupling regime. Near this maximum and in the BEC regime, numerical calculations indicate a clear separation between a temperature scale $T^*$ at which pairing correlations appear and the BKT transition temperature $T_c$ \cite{Scalettar1989}.

\begin{figure*}[ht]
\includegraphics{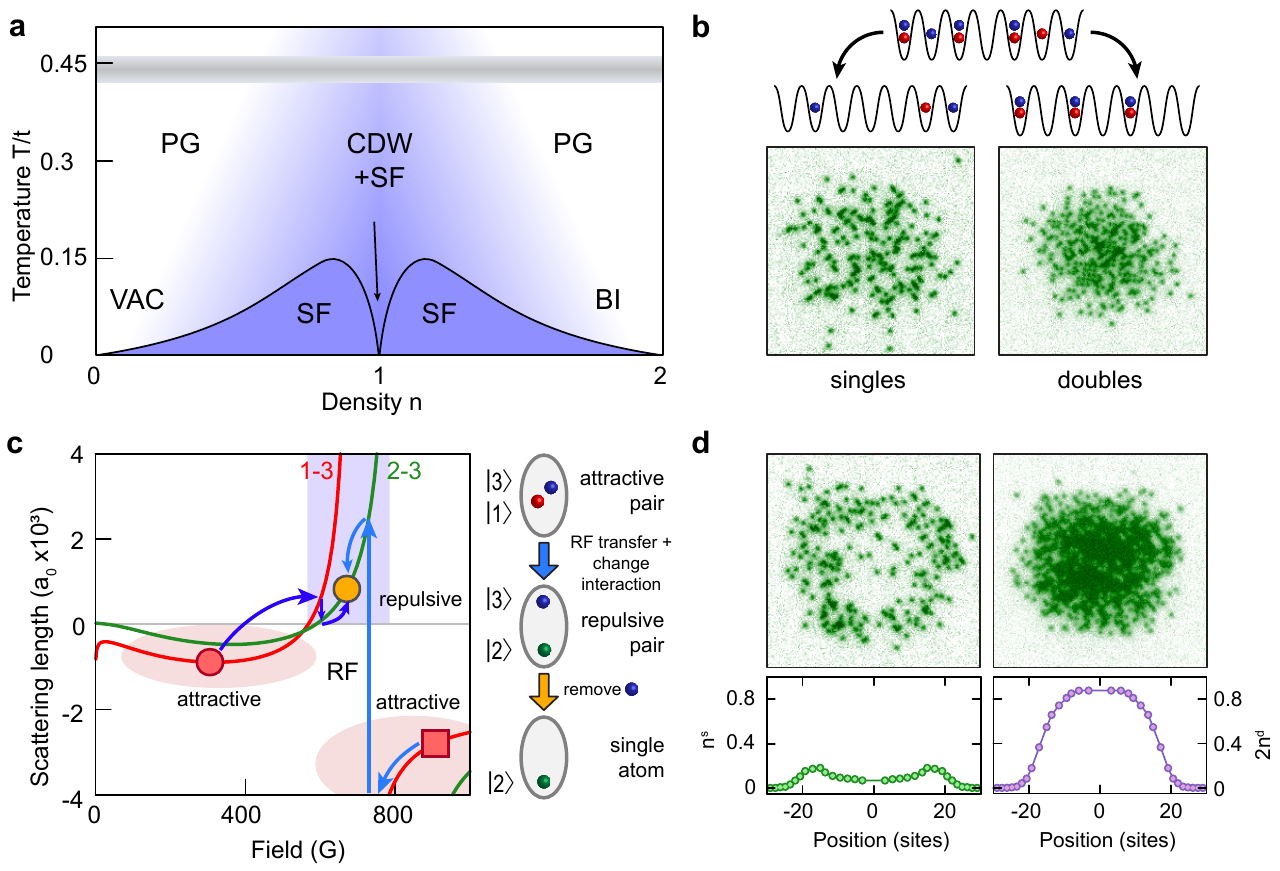}
\caption{\textbf{Experimental scheme for detecting densities and density correlations.} \textbf{a}, Schematic phase diagram of the attractive Hubbard model, indicating pseudogap (PG), superfluid (SF), charge-density-wave (CDW) and band-insulating (BI) regimes. The temperatures achieved in the experiment (gray band) are above the BKT critical temperature but precursor correlations are present in the system. \textbf{b}, Single-shot fluorescence images of the single density $n^s$ (left) and the double density $n^d$ (right). In any single image, we can either detect atoms in the singly occupied sites of the lattice only or in the doubly occupied sites only. \textbf{c}, Overview of magnetic fields and interactions used in the experimental sequence. Lines represent $^6$Li scattering length versus field for a 1-3 mixture (red) and a 2-3 mixture (green) \cite{Zurn2013}. On the upper branch we load the lattice at attractive interactions of $a_{13}=-889~a_0$ (red circle). After freezing the density distribution, we convert 1-3 doublons to 2-3 doublons of which the state $\left|3\right>$ atom is pushed in a regime of repulsive interactions and we are left with only one atom per site (orange circle at $a_{23}=414~a_0$). On the lower branch the lattice is loaded at a scattering length $a_{13}=-2800~a_0$ (red square) and a corresponding doublons detection sequence is performed using the same repulsive interaction (orange circle) for removing one atom in the pair. The diagram to the right illustrates the two steps of the doublon detection: First an interaction dependent transfer and switch to repulsive interactions and then the pushing of one state of the pair which only works well at repulsive interactions. \textbf{d}, Calibration of the efficiency of the scheme to detect doublons in single-site imaging using a band insulating region in the center of the trap. Top, single image of singles (left) and doublons (right). Bottom, azimuthal average of 10 images of singles (left) and doubles (right). In the center, single occupancy is largely suppressed, while the double occupancy exhibits a plateau. Lines are atomic limit fits to the density profiles. \label{fig:schematic}}
\end{figure*}

The phase diagram of the attractive Hubbard model versus filling is shown in Fig.~\ref{fig:schematic}a. At half-filling, the model has a ``hidden" SU(2) pseudo-spin symmetry, in addition to the ordinary SU(2) spin symmetry \cite{Yang1990}. Pseudo-spin rotations on site $\textbf{r}=(r_x,r_y)$ are generated by the charge-density fluctuation operator $\eta^z_{\textbf{r}} = \frac{1}{2}(n_{\textbf{r}} - 1)$ and the pairing operators $\eta_\textbf{r}^{-} = (-1)^{r_x+r_y} c_{\textbf{r},\uparrow}c_{\textbf{r},\downarrow}$ and $\eta^{+}_{\textbf{r}} = (\eta_\textbf{r}^{-})^{\dagger} $. The local pseudo-spin vector is defined by  $(\eta^x_{\textbf{r}}, \eta^y_{\textbf{r}}, \eta^z_{\textbf{r}})$, where $\eta^{\pm}_\textbf{r}=\eta^x_\textbf{r} \pm i\eta^y_\textbf{r}$. The Hubbard Hamiltonian is rotationally invariant under global pseudo-spin rotation at half-filling, leading to checkerboard charge-density-wave (CDW) and superfluid correlations of equal strength. The SU(2) symmetry drives the critical temperature to zero. Away from half-filling, the degeneracy is lifted, and superfluid correlations exhibit quasi-long-range order below a non-zero BKT critical temperature while the charge-density-wave correlations remain short-range. 

Here we perform a site-resolved study of the attractive Hubbard model with a fermionic quantum gas microscope \cite{Haller2015,Edge2015,Omran2015,Parsons2015,Cheuk2015,Yamamoto2016,Brown2016} at temperatures above the BKT transition temperature. We measure the in-trap distributions of  single and double occupancies and observe charge density wave (CDW) correlations near half-filling. This observation allows us to put a lower bound on superfluid correlations in the system. Furthermore, the CDW correlations enable us to perform accurate thermometry on the attractive lattice gas at the superexchange scale. Thermometry at this scale will be important for future work aimed at observing lattice superfluid phases, including homogeneous superfluids in spin-balanced gases and superfluids with spatially-modulated gaps in spin-imbalanced gases \cite{Fulde1964}.

We realize the 2D Fermi-Hubbard model using a degenerate mixture of two hyperfine ground states, $\ket{\uparrow}\equiv\ket{1} $ and $\ket{\downarrow}\equiv\ket{3}$, of $^6$Li in an optical lattice, where $\ket{k}$ labels the $k$th lowest hyperfine ground state. A spin-balanced mixture is obtained by optical evaporation in the vicinity of a broad Feshbach resonance centered at 690~G. After the evaporation, the scattering length is set to $-889~a_0$, where $a_0$ is the Bohr radius, obtained by adjusting a bias magnetic field to 305.4(1)~G. The mixture is prepared in a single layer of an accordion lattice (for details see \cite{Mitra2016}) and subsequently loaded adiabatically into a 2D square lattice of depth $6.2(2)~E_R$ , where $t = h\times 1150(50)$~Hz. The lattice depth is chosen experimentally to maximize the observed CDW correlations at half filling (see Supplementary Information). Here, $E_R= (\pi \hbar)^2 /2ma_{latt}^2 = h\times14.66$~kHz is the recoil energy, where $a_{latt}=1064~\text{nm}/\sqrt{2}$ is the lattice constant. For these parameters, we obtain $U/t=-5.4(3)$ from a bandstructure calculation and a spectroscopic determination of the interaction energy.

We extract the density profiles and correlations in the cloud from site-resolved fluorescence images obtained using quantum gas microscopy techniques (Fig.~\ref{fig:schematic}b). After freezing the density distribution in a deep lattice, we shine near-resonant light on the atoms in a Raman-cooling configuration. Light-assisted collisions eliminate atoms on doubly occupied sites, and we measure the singles density $n^s = n_\uparrow + n_\downarrow - 2 n_\uparrow n_\downarrow$. To gain the full density information, one needs to measure the doubly occupied sites as well. But they are not directly accessible, so we developed a procedure to selectively image doubly occupied sites (Fig.~\ref{fig:schematic}c). After freezing the dynamics, we tag doubly occupied sites using a radiofrequency pulse to transfer atoms in state $\ket{1}$ to state $\ket{2}$ only on these sites, relying on the interaction energy for spectroscopic addressing. We then push atoms in states $\ket{1}$ and $\ket{3}$ out of the lattice with a resonant light pulse. We avoid the loss of state $\ket{2}$ atoms during the resonant light pulse due to light-assisted collisions by an appropriate choice of a positive scattering length for this step. This procedure gives us access to the doublon density $n^d = n_\uparrow n_\downarrow$. We have measured our doublon detection efficiency to be $0.91(1)$ by analyzing the imaging fidelity of band insulating regions in the cloud (Fig.~\ref{fig:schematic}d, see Supplementary Information). The average singles and doubles density profiles obtained from clouds prepared under identical conditions allow us to extract the total density profile $n = n^s + 2 n^d$. For most of our measurements, we adjust the atom number to obtain a total density slightly above $n = 1$ at the center of the trap to obtain a large region in the cloud near half filling.

\begin{figure}
\includegraphics[scale=0.95]{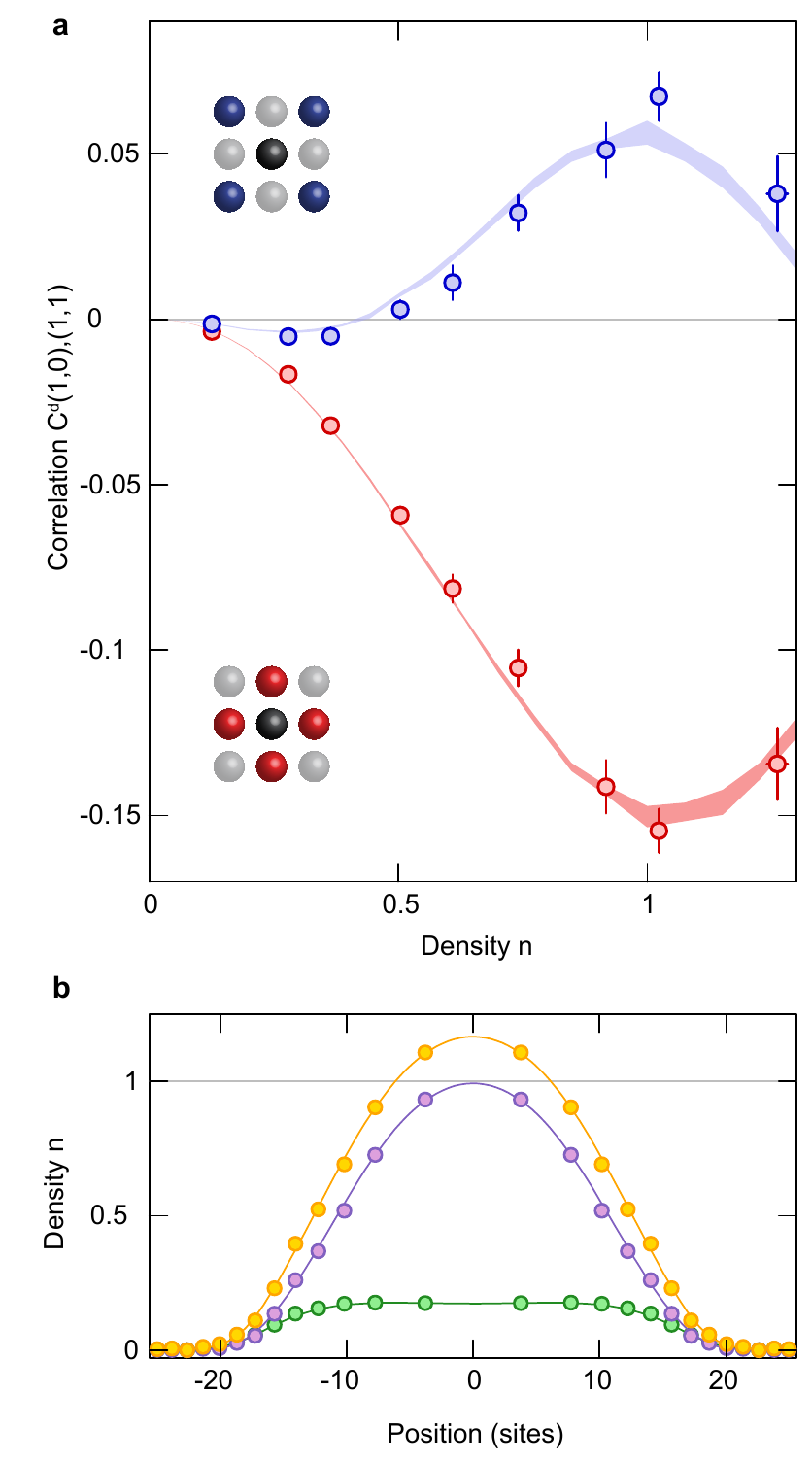}
\caption{\textbf{Observation of charge density wave correlations}. \textbf{a}, Measured nearest-neighbor doublon-doublon correlator $C^d(1,0)$ (red circles) and next-nearest-neighbor correlator $C^d(1,1)$ (blue circles) from an average of 60 repetitions. Error bars s.e.m. DQMC results for $U/t=-5.7$ and $T/t=0.45$ are shown for comparison (bands are s.e.m. of the numerics).  \textbf{b}, Density profiles. Density of singles ($n^s$, green circles), density of doubles ($2n^d$, purple circles) and  total density ($n = n^s + 2n^d$, orange circles). Lines are a simultaneous local density approximation fit of DQMC data to the total and singles densities. For this fit, we fix $U/t$ and $T/t$ at the values above and we obtain from the fit the trap frequency $\omega = 2\pi\cdot\SI{202(5)}{Hz}$ and the central chemical potential $\mu(0) = 0.53(2) U$. \label{fig:cdw_filling}}
\end{figure}

\begin{figure*}
\includegraphics{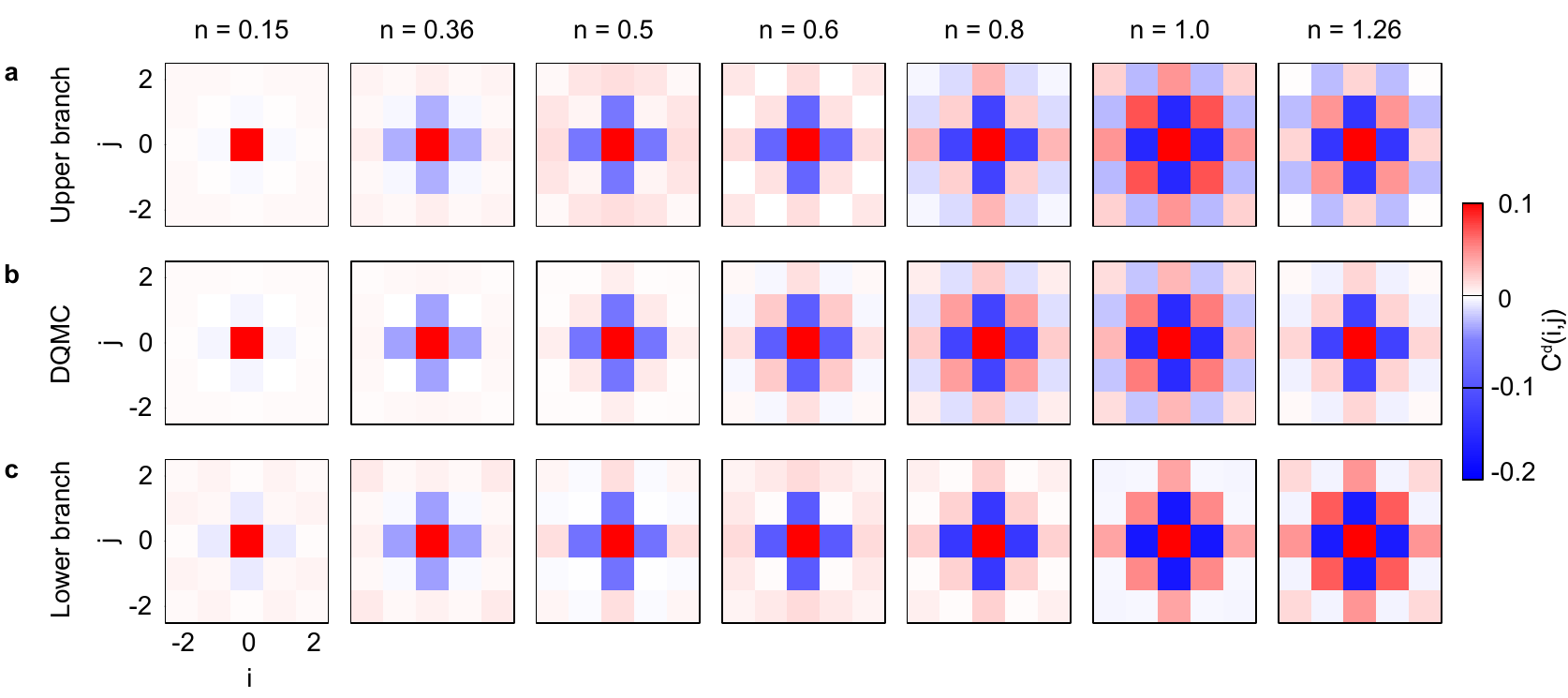}
\caption{\textbf{Doublon-doublon correlation matrices for varying density}. \textbf{a}, Doublon-doublon correlator $C^d(i,j)$ up to three sites for average densities ranging from 0.15 to 1.26 on the upper branch. \textbf{b}, DQMC matrices calculated for the same parameters as in Fig.~\ref{fig:cdw_filling}. \textbf{c}, Doublon-doublon correlator for systems prepared on the lower branch of the Feshbach resonance. Correlator values are averaged over 60 pictures and over symmetric points.  \label{fig:corr_matrices}}
\end{figure*}

Spatial correlations of the $z$-component of the pseudo-spin near half-filling correspond to CDW correlations, characterized by $\langle \eta^z_{\textbf{r}}\eta^z_{\textbf{r}+\textbf{a}}\rangle_c = \langle \eta^z_{\textbf{r}}\eta^z_{\textbf{r}+\textbf{a}}\rangle - \langle \eta^z_{\textbf{r}}\rangle\langle\eta^z_{\textbf{r}+\textbf{a}}\rangle = \frac{1}{4}\langle n_{\textbf{r}}n_{\textbf{r}+\textbf{a}}\rangle_c$ where $\textbf{a}$ is the displacement vector between two lattice sites. We extract a closely related quantity, the doublon-doublon correlator $C^d(\textbf{a}) = 4\langle n^d_{\textbf{r}} n^d_{\textbf{r}+\textbf{a}}\rangle_c$, which also exhibits checkerboard order since most of the atoms in the gas are paired. As the attractive gas is compressible for any filling below unit filling, the local density varies across the harmonic trap. Fig.~\ref{fig:cdw_filling}a shows the doublon-doublon correlator versus density for the nearest-neighbor and the next-nearest-neighbor, obtained by azimuthally averaging the correlations over the trap. The corresponding trap density profile of the gas is depicted in Fig.~\ref{fig:cdw_filling}b. The nearest-neighbor doublon-doublon correlator measured at half-filling is $C^d(1,0) = -0.155(6).$  Since $|C^d(\textbf{a})| \leq \langle\Delta^x_{\textbf{r}} \Delta^x_{\textbf{r}+\textbf{a}}\rangle_c$ for any filling (see Supplementary Information), our measurements constitute a lower bound for superfluid correlations in the system. Here, the $s$-wave superfluid gap operator on site $\textbf{r}$ is defined by $\Delta^x_\textbf{r} = c^{\dagger}_{\textbf{r},\downarrow} c^{\dagger}_{\textbf{r},\uparrow}+c_{\textbf{r},\uparrow} c_{\textbf{r},\downarrow}$.

An interesting feature of the next-nearest-neighbor correlator  $C^d(1,1)$ is that it becomes negative as the average density falls below $\sim 0.4$. This can be understood in the large $U$ limit, where the system can be treated as a gas of hardcore bosons with repulsive nearest-neighbor interactions, leading to negative correlations at distances less than the interparticle spacing. In a recent experiment, a closely related behavior has been observed in the next-nearest-neighbor antiferromagnetic correlations of the $z$-projection of the spin in a spin-imbalanced repulsive Hubbard model \cite{Brown2016}. A particle-hole transformation on operators of $\downarrow$ fermions in the Hubbard Hamiltonian, $c_{\textbf{r},\downarrow}\leftrightarrow (-1)^{r_x+r_y} c^{\dagger}_{\textbf{r},\downarrow}$, changes the sign of the interaction and exchanges the roles of spin-imbalance and doping~\cite{Ho2009}, leading to this symmetry of the correlators between the two experiments.

We theoretically model our system using determinantal Quantum Monte Carlo (DQMC) \cite{Varney2009} in a local density approximation (LDA) and see very good agreement between theory and experiment for the doublon-doublon correlators and density (Fig.~\ref{fig:cdw_filling}). The fits give $U/t=-5.7(2)$ and $T/t =0.45(3)$.  The measured temperature is comparable to our recent measurements in a repulsive gas \cite{Brown2016}. We have also compared to DQMC in the presence of a spatially varying potential to reproduce the largest experimentally observed density gradients to verify that the LDA holds in our system.  

Single-site imaging of doublons allows us to also measure longer range correlations (Fig.~\ref{fig:corr_matrices}). We see doublon-doublon correlations up to two sites on a diagonal shown in the correlation matrices $C^d(i,j)$. We find good agreement with DQMC calculations corresponding to the experimental fillings, calculated using the same parameters as above. The range of the correlation becomes maximal at half filling as expected from theory. 

The previous discussion focused on the upper branch of the Feshbach resonance, used for repulsive Hubbard experiments with lithium \cite{Hart2015,Parsons2016,Boll2016,Brown2016}. Since we are studying an attractive model, we also have the choice of using the lower branch of the resonance, previously used in continuum BEC-BCS crossover experiments \cite{Inguscio2008}. It is not clear \textit{a priori} how the temperature of gases prepared on the two branches would compare. We have explored this question by measuring doublon-doublon correlations in lower-branch gases (Fig.~\ref{fig:corr_matrices}c). The least negative scattering length we can access on this branch is $-2800 a_0$, so we use a lower lattice depth of 4.1(1)~$E_R$ to reach the same value of $U/t$ as on the upper branch. We obtain nearest-neighbor doublon-doublon correlations of -0.172(5) at half filling, suggesting that we reach similar temperatures for this branch. The correlation matrices for the upper and lower branch do not agree very well, especially for larger site separations, which is expected since the on-site interaction energy is comparable to the vertical lattice confinement, leading to higher-band effects which modify the Hamiltonian. A precise determination of the temperature on the upper branch requires taking these effects into account, an endeavor that is outside the scope of this work.

\begin{figure*}
\includegraphics{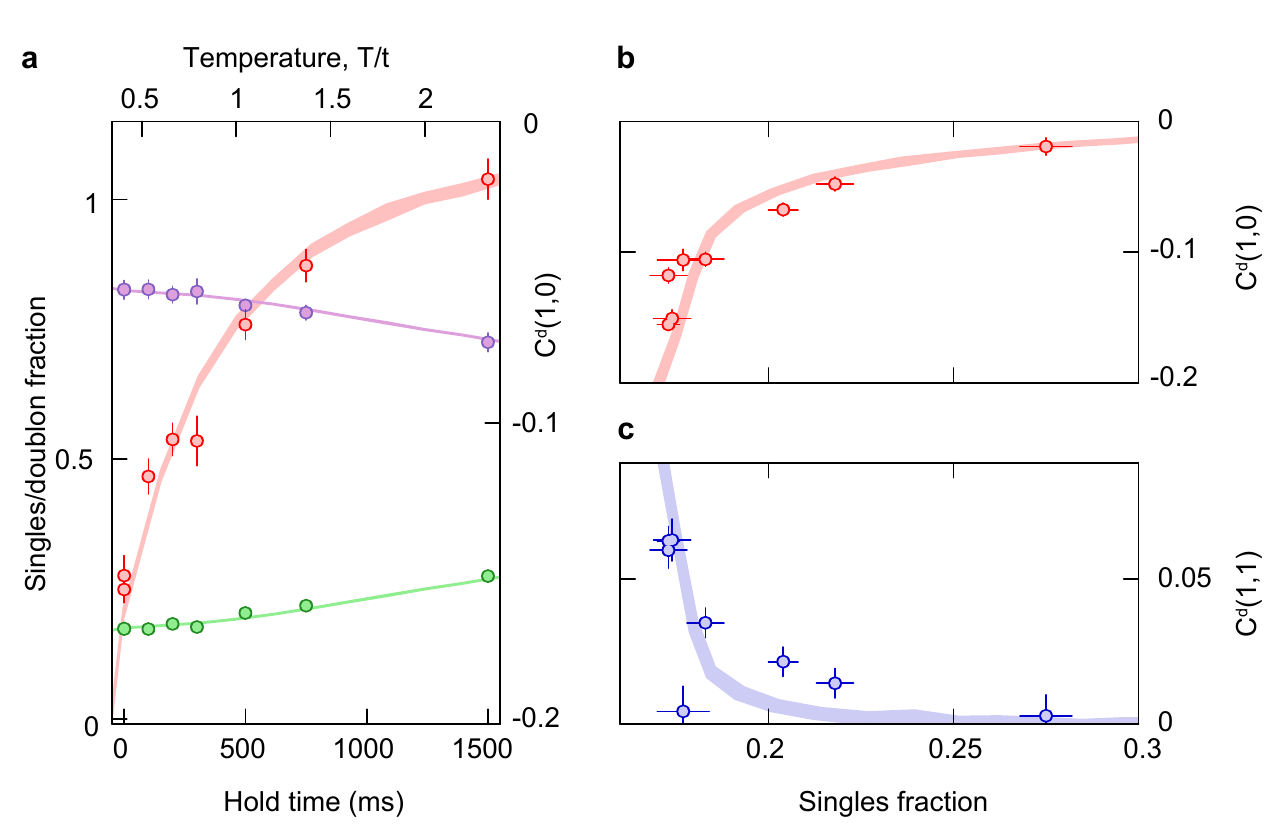}
\caption{\textbf{Thermometry of an attractive Hubbard system.} \textbf{a}, Temperature dependence of pairing and nearest-neighbor doublon-doublon correlations. Shown are the fraction of singles (green circles) and doubles (purple circles) at half-filling on the left axis as a function of hold time in the lattice. On the right axis is the doublon-doublon correlator $C^d(1,0)$ (red circles) at half-filling measured for the same hold times. The upper $x$-axis are temperatures obtained from comparison to DQMC, giving a linear heating rate of 1.3(1) $t/s$. \textbf{b}, $C^d(1,0)$ as a function of singles fraction. Each data point corresponds to a single hold time. \textbf{c}, The next-nearest neighbor doublon-doublon correlator $C^d(1,1)$ as function of singles fraction.  Bands correspond to DQMC results for $U/t=-5.7$ and half filling in all of the above. Each point is averaged over 60 pictures. Error bars and bands are s.e.m. \label{fig:cdw_temperature}}
\end{figure*}


Both pairing and CDW correlations may be used as thermometers in an attractive Hubbard gas. To investigate their sensitivity as thermometers, we heated the upper branch system in a controlled fashion by holding it for variable times in the lattice before imaging. Technical noise leads to a linear increase in the temperature of the system. We observe a slight reduction in doublon fraction in a region of half filling for long hold times (Fig.~\ref{fig:cdw_temperature}a), while the doublon-doublon correlators $C^d(1,0)$ and $C^d(1,1)$ show a significant change during the same time. These observations illustrate that the doublon fraction is a good thermometer for temperatures on the order of $U$, where CDW correlations are small, while the doublon-doublon correlator is a more sensitive thermometer for temperatures on the order of the exchange $4t^2/U$. Fig.~\ref{fig:cdw_temperature}b,c  shows the nearest-neighbor and next-nearest-neighbor doublon-doublon correlators versus the singles fraction at half-filling. Plotting the data this way allows a temperature-independent comparison to DQMC with a single free-parameter ($U/t$), and we see good agreement for $U/t=-5.7$.


In conclusion, we have performed quantum gas microscopy on an attractive atomic Fermi-Hubbard system. We observed CDW correlations and studied their dependence on the lattice filling and temperature. We have shown that these correlations serve as an excellent thermometer in the low temperature regime that will be useful in the quest to reduce the temperature of attractive lattice systems to observe long-range Hubbard superfluids. Above the critical temperature, attractive Hubbard systems will allow experimental exploration of the physics of the pseudogap regime, while at lower temperatures, they will enable studies of the BKT transition in a lattice system. The superfluid correlations inferred in this experiment may be probed more directly using collective excitations of the superfluid order parameter that couple to the conjugate CDW order parameter \cite{Demler1996}. Another interesting direction for future work is the study of the spin-imbalanced attractive Hubbard model in 2D, where Fermi surface nesting due to the lattice increases the stability of Fulde-Ferrell-Larkin-Ovchinnikov (FFLO) superfluids \cite{Moreo2007,Loh2010,Kim2012,Gukelberger2016}. Unlike spin-balanced gases investigated in this work, DQMC suffers a fermion ``sign" problem in simulating the spin-imbalanced attractive Hubbard model, making theoretical predictions difficult at low temperatures.

\begin{acknowledgments}
This work was supported by the NSF (grant no. DMR-1607277), the David and Lucile Packard Foundation (grant no. 2016-65128), and the AFOSR Young Investigator Research Program (grant no. FA9550-16-1-0269). W.S.B. was supported by an Alfred P. Sloan Foundation fellowship. P.T.B. was supported by the DoD through the NDSEG Fellowship Program. 
\end{acknowledgments}

%


\pagebreak
\clearpage
\setcounter{equation}{0}
\setcounter{figure}{0}

\renewcommand{\theparagraph}{\bf}
\renewcommand{\thefigure}{S\arabic{figure}}
\renewcommand{\theequation}{S\arabic{equation}}

\onecolumngrid
\flushbottom

\section{\Large Methods}

{\bf Preparation of an attractive Fermi gas in an optical lattice} The experimental setup is described in detail in the supplement of ref.~\cite{Brown2016}. We realize the Fermi-Hubbard model using a spin-balanced degenerate mixture of two Zeeman states ($\ket{1}=\ket{\uparrow}$ and $\ket{3}=\ket{\downarrow}$, numbered up from the lowest energy) in the ground state hyperfine manifold of $^6$Li in an optical lattice. 

To create the sample we load a magneto-optical trap (MOT) from a Zeeman slower, then use a compressed MOT stage before loading into a $\approx$ \SI{1}{\milli \kelvin} deep optical trap and evaporating near the \SI{690}{G} Feshbach resonance. For preparing a gas on the upper branch of the resonance, we stop the evaporation before Feshbach molecules form and transfer the atoms to a highly anisotropic `light sheet' trap where the gas undergoes further evaporation to degeneracy at \SI{305.4(1)}{G}. At this field, the scattering length is $a_s = -889 a_0$. Next we transfer the gas into the final trapping geometry where a \SI{1070}{\nm} beam provides radial confinement and a \SI{532}{\nano\meter} accordion lattice with trapping frequency $\omega_z = 2 \pi \times$ \SI{21(1)}{\kilo\hertz} provides axial confinement (for further details see \cite{Mitra2016}). The spin populations are balanced to within $2.1(9)\%$.  We then load the gas into a 2D square lattice with a \SI{25}{\milli\second} long ramp to varying depths from $4-7.5~E_r$. 

For experiments on the lower branch of the resonance, we perform the entire evaporation at the Feshbach resonance. Three-body collisions lead to population of the molecular branch. Before loading into the lattice, the bias field is ramped to 907~G where the scattering length is -2800~$a_0$.

{\bf Calibration of Hubbard parameters} We use a non-separable 2D square lattice with a \SI{752}{\nano\meter} spacing formed by four interfering passes of a single vertically polarized beam \cite{Brown2016}. We calibrate our lattice depth by measuring the frequencies of the three $d$ bands in a deep lattice using lattice depth modulation, and compare these with a 2D band structure calculation. The inferred depth of the lattice at which our measurements are performed is \SI{6.2(2)}~$E_r$, where $E_r = \SI{14.66}{\kilo\hertz}$. From that we obtain nearest-neighbor tunneling values $t_x = 1200$ Hz, $t_y = 1110$ Hz ($t_x/t_y = 1.08$). The reduction of the lattice depth across the cloud due to the gaussian profile of the lattice beams leads to an increase in the tunneling by \SI{10}{\%} at the edge of the cloud compared to the central value. We also calculate a non-zero but negligible diagonal tunneling $t_{11} = \SI{42}{Hz} = 0.04 t_x$, due to the non-separability of the lattice.

We measure the interaction energy $U$ at the lattice loading field of 305~G using radio frequency spectroscopy. We transfer atoms from state $\ket{1}$ to $\ket{2}$ and resolve the frequency shift between singly and doubly occupied sites. We determine $U_{13} = \delta U \frac{a_{13}}{a_{13}-a_{23}}$, where $\delta U$ is the measured difference between the singles and doublon peaks and $a_{13}$ ($a_{23}$) is the scattering length at the spectroscopy field for a 1-3 (2-3) mixture. Taking into account a correction due to a significant final state interaction, we obtain $U_{13}=6.6(3)$~kHz. The experimentally measured value agrees with the value determined from band structure calculations of 5.9(1)~kHz including higher band corrections to within 10\% \cite{Idziaszek2005}.

\begin{figure*}
\includegraphics{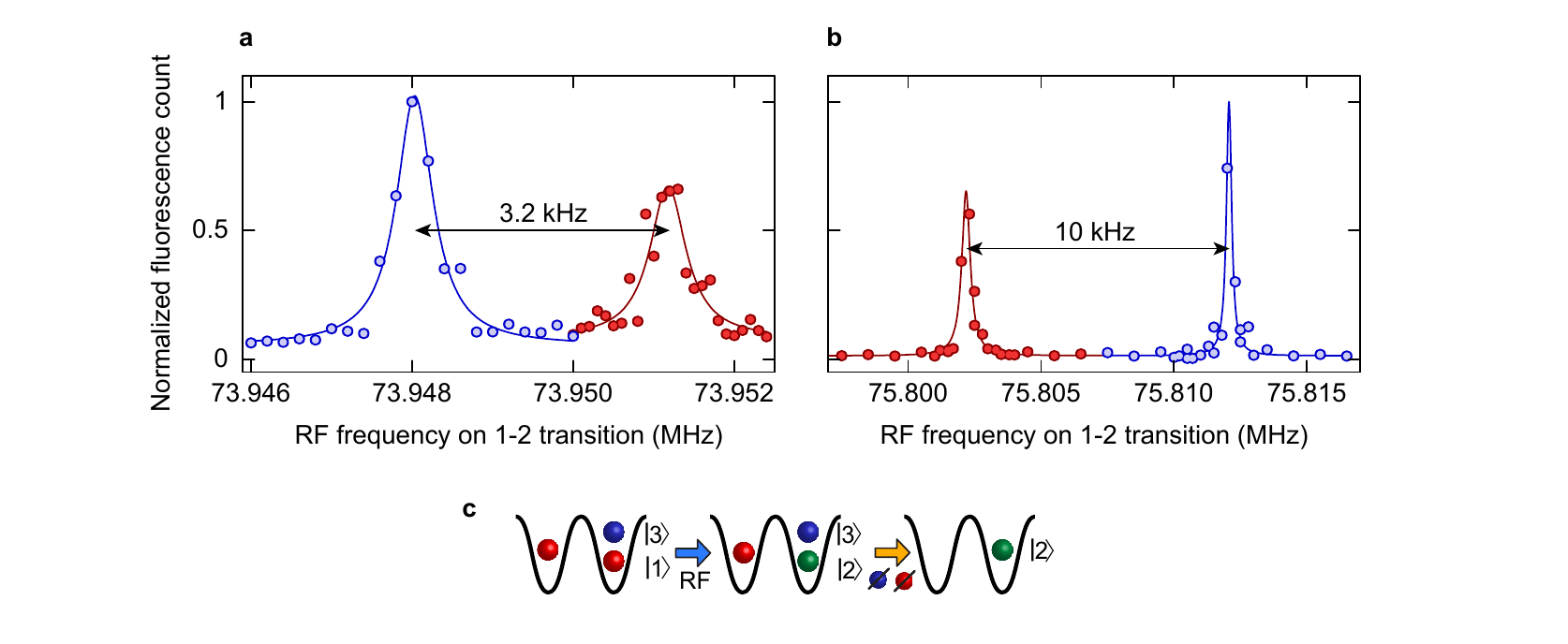}
\caption{ \textbf{Radio-frequency spectra of atoms in the lattice}. $\ket{1}$-$\ket{2}$ radio-frequency spectra showing excitation of unpaired $\ket{1}$ atoms (red) and paired $\ket{1}$ atoms (blue). Lines are Lorentzian fits to the data. \textbf{a}, RF spectrum at a field of 305.4~G and a lattice depth of $6.2(2)~E_R$. The doublon transition is lowered in energy relative to the singles transition due to attractive interactions. \textbf{b}, RF spectrum at 594~G and a lattice depth of $55(1)~E_R$. The doublon transition is shifted upwards from the singles transition by 10~kHz due to repulsive interactions. \textbf{c}, Schematic of our doublon imaging process. The RF pulse is applied at 594~G where both initial $\ket{1}$-$\ket{3}$ and final $\ket{2}$-$\ket{3}$ states are repulsively interacting. We use a Landau-Zener sweep that is centered on the doublon transition (blue peak in \textbf{b}), selectively exciting paired $\ket{1}$ atoms. Subsequent resonant pulses remove atoms in states $\ket{1}$ and $\ket{3}$, leaving only $\ket{2}$ atoms behind for fluorescence imaging.}
\label{fig:doublon-imaging}
\end{figure*}

{\bf Imaging of doublons} In the usual scheme of Raman imaging, a singly occupied lattice site produces a fluorescence signal while atoms in a doubly occupied site are lost due to light-assisted collisions. To measure density correlations between doubly-occupied sites, we have developed a new detection scheme. After the gas is loaded adiabatically into the optical lattice, we pin the atoms by increasing the lattice depth to 55(1)~$E_R$ in 100~$\mu s$ where tunnelling dynamics get frozen. For gases on the upper branch, we adiabatically ramp the field to 594~G where we perform an interaction resolved Landau-Zener sweep (Fig.~\ref{fig:doublon-imaging}) to selectively transfer 1-3 doublons to 2-3 doublons while not effecting the $\ket{1}$ singles. Finally we ramp to a field of 641~G where the 2-3 scattering length is 414~$a_0$. We apply resonant pushing pulses of 30~$\mu s$ durations to remove $\ket{1}$ and $\ket{3}$ atoms, leaving behind with only $\ket{2}$ atoms on sites that originally had doublons. The large relative wavefunction of the atoms on a site due to repulsive interactions and a relatively weak vertical confinement significantly reduces the probability of light-assisted collisions during the resonant pulse. On the other hand, in order to get the singles images, we apply no RF or resonant pushing pulses.

For gases on the lower branch, after pinning the atoms, we ramp the field to 725~G where we perform a Landau-Zener sweep on an interaction-resolved transition to the 2-3 upper branch. In order to convert the 2-3 doubles to $\ket{2}$-singles with high efficiency, we ramp to 641~G where we apply the resonant pulses to remove $\ket{1}$ and $\ket{3}$ atoms. This ramp is done quickly (within $\sim$800~$\mu s$) to avoid atom losses due to the crossing of a narrow 2-3 Feshbach resonance around 714~G.

We image the final atom distribution by increasing the lattice depth to 2500 $E_r$ within $250~\mu$s, ramping up the light sheet to provide axial confinement, and then collecting fluorescence photons during Raman cooling.

To measure the efficiency of imaging doublons, we prepare a band insulator where the filling in the trap center is saturated at two atoms per site. We perform the above-mentioned process to image only the doublons. We measure a combined fidelity (including RF-transfer efficiency and pushing efficiency) of $1-\epsilon_d= 90.9(5)\%$ of imaging doublons, leading to an underestimation of our doublon-doublon correlator by $(1-\epsilon_d)^2=0.83$ which we correct for. We have also performed the same detection procedure on a Mott insulator, where we expect unit occupancy on the lattice sites. This allows us to extract the probability that a single atom in state $\ket{1}$ would get transferred to $\ket{2}$ and give a false positive signal of the presence of a doublon. We measure the probability of this process to be only 2.3(3)$\%$.

{\bf Raman imaging and reconstruction} We perform Raman imaging for \SI{1200}{\milli \second} and collect approximately $1000$ photons per atom. For more details, see \cite{Brown2016}. We estimate fidelity errors due to Raman imaging imperfections by taking 40 consecutive images of the same atom cloud and determine the shot-to-shot differences. This leads to a hopping rate during one picture of \SI{0.4(1)}{\%} and a loss rate of \SI{1.9(2)}{\%}. In addition, while holding the atoms in a deep lattice for doublon detection, we lose \SI{2(1)}{\%} of the atoms, leading to a net detection efficiency of $\approx \SI{95}{\%}$. The densities that we obtain are corrected for the above detection efficiency.

{\bf Lower bound on superfluid correlations} 
The full Hubbard Hamiltonian can be written as 
\begin{equation} \label{eq:H}
{\cal {H}}=-t\sum_{\langle \textbf{rr}^\prime\rangle,\sigma} \left(c^{\dag}_{\textbf{r},\sigma}c_{\textbf{r}^{\prime},\sigma}+c^{\dag}_{\textbf{r}^{\prime},\sigma}c_{\textbf{r},\sigma}\right) + U \sum_\textbf{r} n_{\textbf{r},\uparrow}n_{\textbf{r},\downarrow}-\mu\sum_{\textbf{r},\sigma} n_{\textbf{r},\sigma}-h\sum_{\textbf{r}}(n_{\textbf{r},\uparrow}-n_{\textbf{r},\downarrow})
\end{equation}
where $c^{\dag}_{\textbf{r},\sigma}$ is the creation operator for a fermion with spin $\sigma$ on site $\textbf{r}$, $n_{\textbf{r},\sigma} = c^{\dag}_{\textbf{r},\sigma}c_{\textbf{r}\sigma}$, $t$ is the matrix element for tunneling between lattice sites, $U$ is the strength of the on-site interaction ($U < 0$ for the attractive model), $\mu$ is a spin-independent chemical potential and $h$ is an effective Zeeman field in the presence of spin-imbalance. For the purpose of this paper, $h=0$, since we only work with a spin-balanced system. Consider first the SU(2) spin symmetry of the problem for $h=0$. The vector spin operator on site $\bf{r}$ is given by $(S^x_{\bf{r}},S^y_{\bf{r}},S^z_{\bf{r}})$. In terms of fermionic creation and annihilation operators, the generators of spin rotations are
\begin{eqnarray}
S^-_{\mathbf{r}} &=&  c_{\mathbf{r},\downarrow}^{\dagger}c_{\mathbf{r},\uparrow} \nonumber\\
S^+_{\mathbf{r}} = (S^-_{\mathbf{r}})^\dagger &=& c_{\mathbf{r},\uparrow}^{\dagger}c_{\mathbf{r},\downarrow}\nonumber \\
S^z_{\mathbf{r}} = \frac{1}{2}( c_{\mathbf{r},\uparrow}^{\dagger}c_{\mathbf{r},\uparrow}- c_{\mathbf{r},\downarrow}^{\dagger}c_{\mathbf{r},\downarrow})&=&\frac{1}{2}(n_{\mathbf{r},\uparrow}-n_{\mathbf{r},\downarrow})
\end{eqnarray}

\begin{figure}[h]
\includegraphics{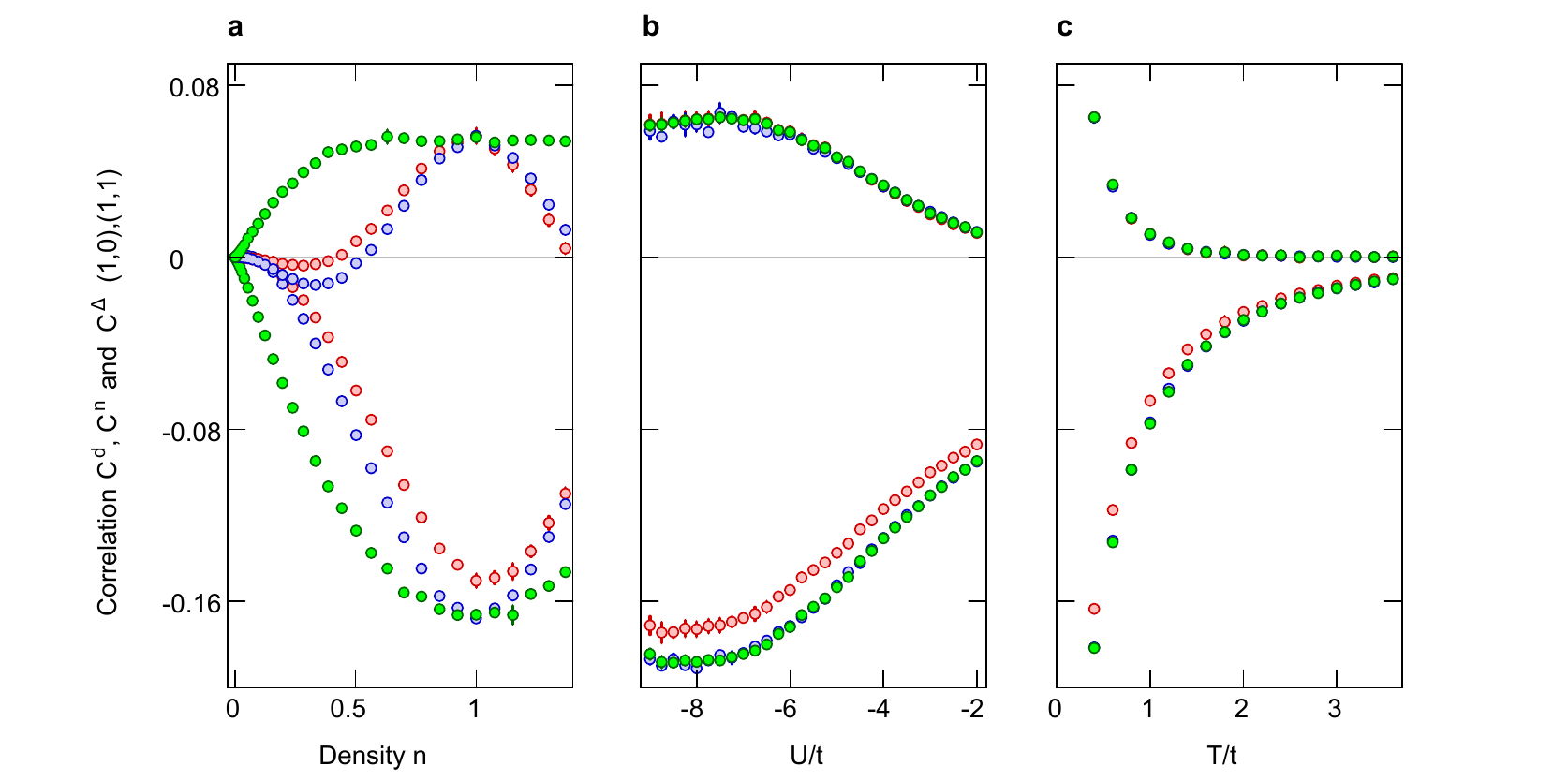}
\caption{ \textbf{DQMC simulations of pairing and CDW correlators}. \textbf{a}, DQMC simulations for the nearest neighbor (1,0) and next-nearest neighbor (1,1) correlations for doublon-doublon ($C^d(\textbf{a})$, red), density-density ($C^n(\textbf{a})$, blue) and gap-gap ($(-1)^{a_x+a_y}C^\Delta(\textbf{a})$, green) as a function of density for $U/t=-5.7$ and $T=0.45t$. \textbf{b}, Same quantities as a function of $U/t$ at half filling ($n=1$) and $T=0.45t$. \textbf{c}, Same quantities as a function of $T/t$ at half filling and for $U/t=-5.7$. Note that for both (1,0) and (1,1) correlators, $\abs{C^\Delta} \geq \abs{C^d}$.}
\label{fig:dqmc}
\end{figure}

These operators obey the usual commutation relations 
\begin{equation}
[S^z_{\mathbf{r}},S^{\pm}_{\mathbf{r}}] = \pm S^{\pm}_{\mathbf{r}} , \quad [S^+_{\mathbf{r}},S^-_{\mathbf{r}}] = 2S^z_{\mathbf{r}}
\end{equation}
In the absence of an effective Zeeman field, the operators $S^z =  \sum_r S^z_{\mathbf{r}}$ and $S^{\pm} =  \sum_r S^{\pm}_{\mathbf{r}}$ satisfy
\begin{equation} \label{eq:comm1}
[{\cal{H}},S^{\pm}] = [{\cal{H}},S^z] = 0
\end{equation} 
implying the SU(2)  spin-symmetry of the Hubbard model. In other words, the Hamiltonian is invariant under a global rotation of the spin degree of freedom. To demonstrate the other ``hidden" symmetry of the Hubbard model, we define a new set of generators for rotations of the ``pseudo-spin" on a site given by
\begin{eqnarray}\label{eq:etas}
\eta_\textbf{r}^{-} &=& (-1)^{r_x+r_y} c_{\textbf{r},\uparrow}c_{\textbf{r},\downarrow}\nonumber\\
\eta_\textbf{r}^+ &=& (\eta_\textbf{r}^-)^\dagger\nonumber\\
\eta^z_{\textbf{r}} &=& \frac{1}{2}(n_{\textbf{r}} - 1)
\end{eqnarray}
It can be easily verified \cite{Zhang1990} that they obey the following commutation relations
\begin{equation}
[\eta^z_\textbf{r},\eta^{\pm}_\textbf{r}]=\pm\eta^{\pm}_\textbf{r}, \quad [\eta^+_\textbf{r},\eta^-_\textbf{r}]=2\eta^z_\textbf{r}.
\end{equation}
The operators $\eta^z =  \sum_r \eta^z_{\mathbf{r}}$ and $\eta^{\pm} =  \sum_r \eta^{\pm}_{\mathbf{r}}$ satisfy
\begin{equation} \label{eq:comm2}
[{\cal{H}},\eta ^\pm] = \pm (U-2\mu)\eta ^{\pm}, \quad [{\cal{H}},\eta ^z] = 0.
\end{equation}
regardless of the sign of the interaction $U$. Exactly at half-filling, $\mu=U/2$ and all the pseudospin generators in Eq.~\ref{eq:comm2} commute with $\cal{H}$ similar to the spin generators in Eq.~\ref{eq:comm1}, meaning that the Hamiltonian is invariant under global rotations of the pseudo-spin.

The pseudo-spin on a site can be visualized on the Bloch sphere like a regular spin. If it points up (down) the site contains a doublon (hole), while if it lies in the equatorial plane, the site contains an equal superposition of a doublon and hole with a complex relative phase determined by the azimuthal angle. In the limit of large attractive interactions, the spin-balanced Hubbard Hamiltonian can be approximated as a Heisenberg Hamiltonian with antiferromagnetic interactions between the pseudo-spins, leading to charge-density-wave and superfluid correlations corresponding to $z$ and $x,y$ antiferromagnetic pseudo-spin correlations respectively. At half-filling, the pseudo-spin rotational symmetry implies
\begin{equation}\label{eq:equality}
\avg{\eta^z_{\mathbf{r}}\eta^z_{\mathbf{r+a}}}_c=\avg{\eta^x_{\mathbf{r}}\eta^x_{\mathbf{r+a}}}_c
\end{equation}
where $\eta^x_\textbf{r} = (\eta^+_\textbf{r} + \eta^-_\textbf{r})/2$. Defining the s-wave pairing operator in the $x$ direction as
\begin{equation} 
\Delta_\textbf{r}^{x} = c_{\textbf{r},\uparrow}c_{\textbf{r},\downarrow}+c^\dagger_{\textbf{r},\downarrow}c^\dagger_{\textbf{r},\uparrow}
\end{equation}
one concludes that the density correlations $C^n(\textbf{a})=\avg{n_\textbf{r}n_\textbf{r+a}}_c = 4\avg{\eta^z_{\textbf{r}}\eta^z_{\textbf{r+a}}}_c$ and pairing correlations $C^\Delta(\textbf{a})=\avg{\Delta^x_\textbf{r}\Delta^x_\textbf{r+a}}_c = 4\avg{\eta^x_{\textbf{r}}\eta^x_{\textbf{r+a}}}_c$ are equal in magnitude at half-filling.

Deviation from half-filling introduces an effective Zeeman field along $z$ that couples to the pseudo-spin. This leads to canted antiferromagnetic pseudo-spin correlations, with stronger correlations in the direction orthogonal to the field, i.e. the superfluid correlations become stronger than the density correlations. Therefore, measurement of charge-density-wave density correlations in the attractive Hubbard model provides a lower bound on the superfluid correlations at any filling. In our experiment we measure a more accessible quantity, the doublon correlator $C^d(\textbf{a}) = 4 \avg{n^d_\textbf{r}n^d_\textbf{r+a}}_c$. The doublon and density correlators become equal in the limit of low temperatures and large interactions. However, the doublon correlator $C^d$ still provides a bound on the superfluid correlator $C^\Delta$ as we have verified numerically for the entire range of interactions, temperatures and fillings studied in the experiment (Fig.~\ref{fig:dqmc}).

\begin{figure}
\includegraphics{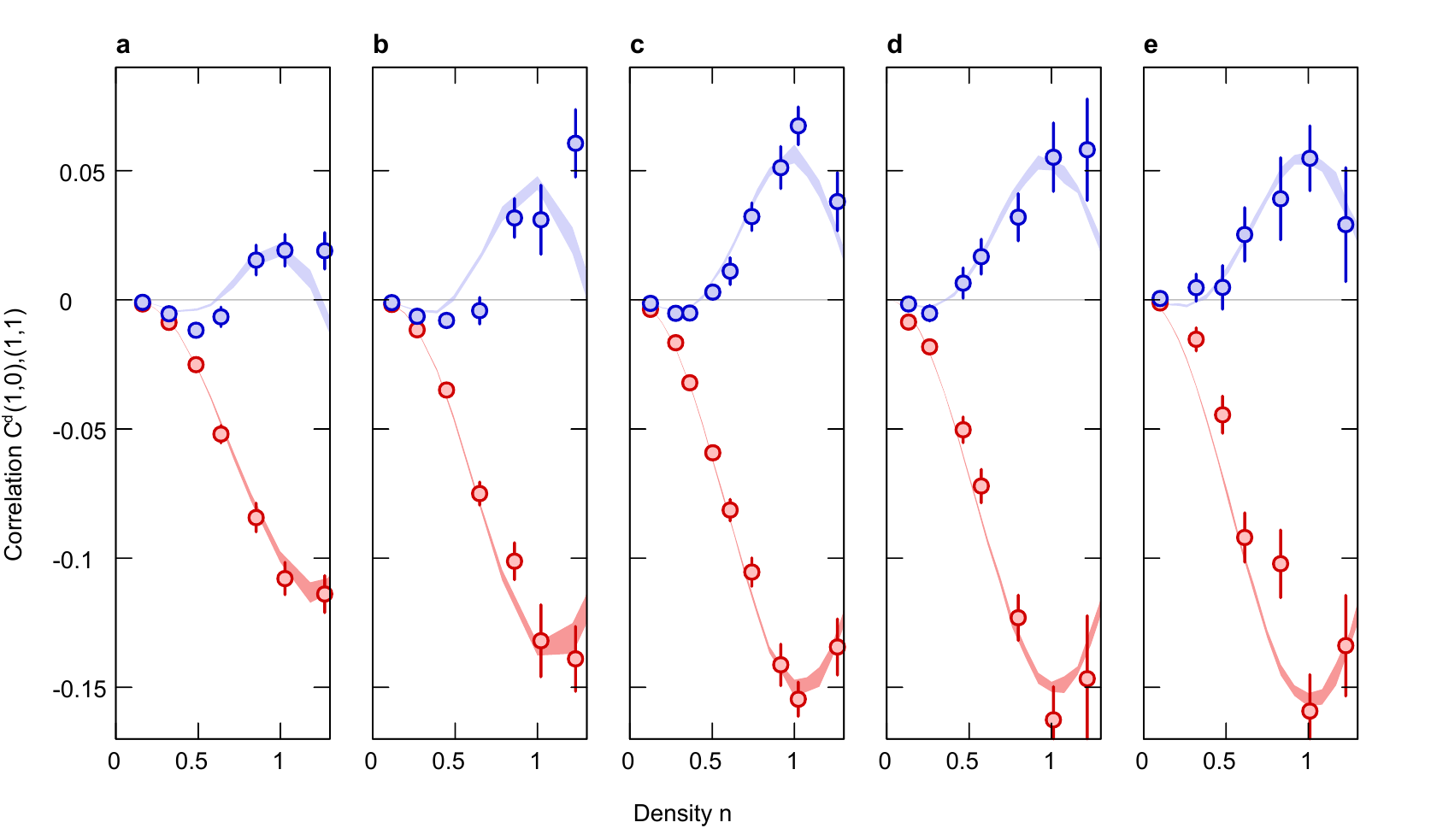}
\caption{\textbf{Nearest-neighbor and next-nearest neighbor doublon-doublon correlators vs. filling as a function of lattice depth}. $C^d(1,0)$ (red circles) and $C^d(1,1)$ (blue circles) obtained from different parts of the trap where filling varies from 0 to slightly above half-filling. \textbf{a}, for lattice depth of 4.1(1)~$E_R$, \textbf{b}, 5.5(1)~$E_R$, \textbf{c}, 6.2(2)~$E_R$, \textbf{d}, 6.7(2)~$E_R$, \textbf{e}, 7.4(2)~$E_R$. Shaded curves are DQMC fits for the data versus filling with $U/t$ and $T/t$ as fit parameters. \textbf{a}, $U/t = -2.75, T=0.4t$, \textbf{b}, $U/t = -4.5, T=0.4t$, \textbf{c}, $U/t = -5.7, T=0.45t$, \textbf{d}, $U/t = -6.75, T=0.5t$, \textbf{e}, $U/t = -7.5, T=0.5t$. The width of the shaded band is the s.e.m of the simulation data.   \label{fig:corr_U}}
\end{figure}

{\bf Relationship to earlier work on repulsive spin-imbalanced Hubbard systems} On a bipartite lattice, one can define a particle-hole transformation given by following mapping of the annihilation operators
\begin{equation}
c_{\mathbf{r},\downarrow} \leftrightarrow (-1)^{r_x+r_y}c^\dagger_{\mathbf{r},\downarrow}, \quad c_{\mathbf{r},\uparrow}\leftrightarrow c_{\mathbf{r},\uparrow}.
\end{equation}
Under this transformation, $n_{\uparrow}\leftrightarrow n_{\uparrow}$ and $n_{\downarrow}\leftrightarrow 1-n_{\downarrow}$ and the Hubbard Hamiltonian (\ref{eq:H}) transforms to (up to a constant energy offset)
\begin{equation}
{\cal{H}}\leftrightarrow -t\sum_{\langle \textbf{rr}^\prime\rangle,\sigma} \left(c^{\dag}_{\textbf{r},\sigma}c_{\textbf{r}^{\prime},\sigma}+c^{\dag}_{\textbf{r}^{\prime},\sigma}c_{\textbf{r},\sigma}\right) - U \sum_\textbf{r} n_{\textbf{r},\uparrow}n_{\textbf{r},\downarrow}-\mu'\sum_{\textbf{r},\sigma} n_{\textbf{r},\sigma}-h'\sum_{\textbf{r}}(n_{\textbf{r},\uparrow}-n_{\textbf{r},\downarrow})
\end{equation}
where, $\mu'=h-U/2$ and $h'=\mu-U/2$. In addition, the pseudo-spin rotation generators map on to the regular spin rotation generators under this transformation
\begin{equation}
\eta^{\pm} \leftrightarrow S^{\pm}, \quad \eta^z \leftrightarrow S^z
\end{equation} 
In other words, this transformation provides a mapping between the attractive and the repulsive Hubbard models, with the roles of doping and spin-imbalance interchanged \cite{Ho2009}. The doped attractive Hubbard system studied in this work maps onto a spin-imbalanced repulsive Hubbard system. Therefore, the $z$-spin correlations measured in our previous work \cite{Brown2016} are closely related to the CDW correlations measured in this work.

{\bf Dependence of CDW  correlations on the interactions} We studied the dependence of the CDW correlations on $U/t$. The system was prepared on the upper branch at a scattering length of $-889~a_0$ and $U/t$ was varied by changing the lattice depth. Fig.~\ref{fig:corr_U} shows the nearest-neighbor and next-nearest-neighbor correlator vs. density for different lattice depths. The data is compared to DQMC results with $U/t$ and $T/t$ used as fit parameters. A slight increase in the temperature is observed for larger lattice depths. The measured correlations initially increase as the lattice depth is increased, but do not show significant variation for lattice depths between $6.2~E_R$ and $7.4~E_R$. At larger depths, it is experimentally difficult to obtain a half-filling condition near the center of the trap because of the increasing radial confinement from the lattice beams.

{\bf Corrections to the Hubbard model} We investigated the importance of corrections to the Hubbard model in our analysis of the experimental data. For this purpose, two effects were taken into account: 

\begin{itemize}
\item{{\bf Effects of next-nearest-neighbor tunneling} From a tight-binding calculation, we obtain for a lattice depth of $6.2~E_R$ (used for upper branch experiments) tunnel couplings $t_{11}/t\sim 4\%$ and $t_{20}/t\sim-10\%$, where $t_{ij}$ is the tunneling matrix element to the (1,1) and (2,0) neighboring site, respectively. We studied the effect of including these terms in the Hamiltonian on the CDW correlations (Fig.~\ref{fig:Hubb_corrections}) and found it to be negligible within our experimental errors.}

\begin{figure}[t]
\includegraphics{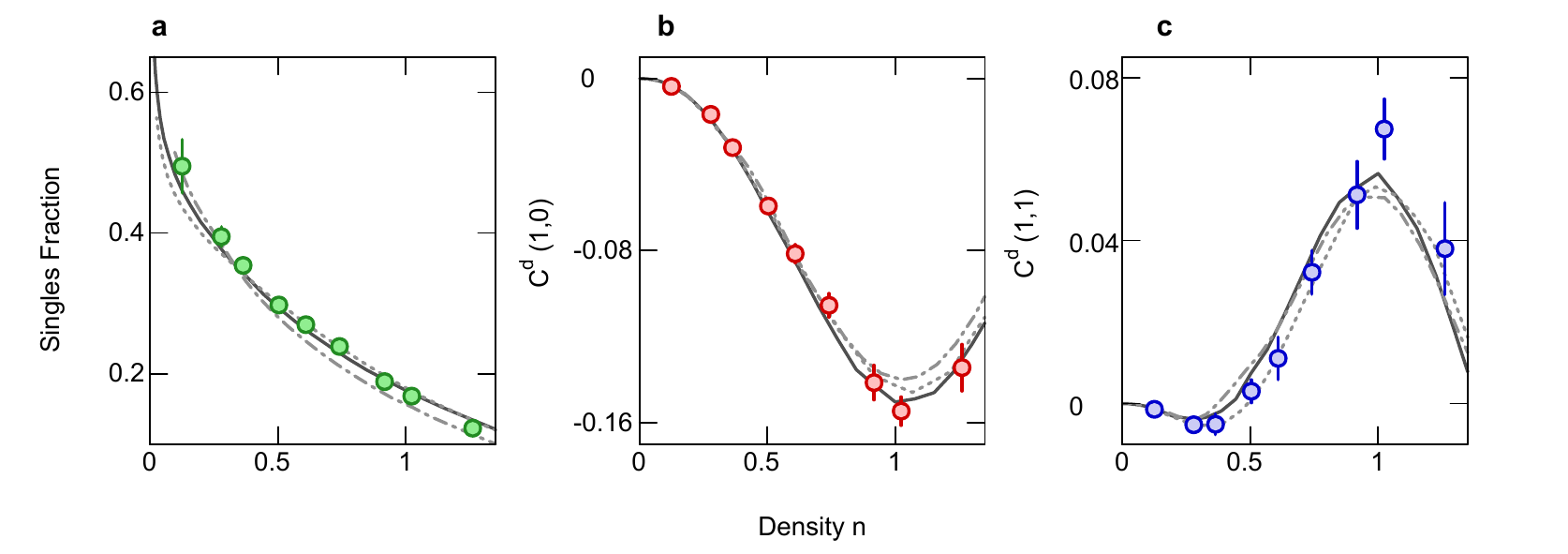}
\caption{ \textbf{Corrections to the Hubbard Model} Numerical study of effect of corrections to the Hubbard model. \textbf{a}. Experimental singles fraction (green circles) as a function of filling for a 6.2(2)~$E_R$ lattice. \textbf{b}. Experimental nearest neighbor doublon-doublon correlator $C^d(1,0)$ (red circles) versus filling. \textbf{c}. Experimental next-nearest neighbor   doublon-doublon correlator $C^d(1,1)$ (blue circles) versus filling. In all of the above, black line represents DQMC Hubbard model simulation results for $U/t=-5.7$ and $T=0.45t$. Gray dotted line represents DQMC simulation results for the above $U/t$ and $T/t$ but including longer-range tunneling terms with $t_{11}/t=4\%$ and $t_{20}/t=-10\%$. Gray dashed line represents NLCE simulation results with density-dependent hopping at the same $U/t$ and $T/t$ for $t_n/t=-8\%$. \label{fig:Hubb_corrections}}
\end{figure}

\item{{\bf Density-dependent tunneling} The simple Hubbard model we use in this paper assumes that the tunneling matrix element for an atom is independent of the occupancy of the site it is tunneling to. This approximation starts to break down as the scattering length $a_s$ becomes large, and it may be expected that this effect is important in our system since most of the atoms are in pairs. The density-dependent hopping can be expressed in a tight-binding model as \cite{Jurgensen2014}
\begin{equation}
t_n = -\frac{4\pi\hbar^2 a_s}{m}\int{d^3rw^*(\textbf{r}-\textbf{d})w^*(\textbf{r})w^2(\textbf{r})}
\end{equation}
where $w(\textbf{r})$ is the Wannier function of an atom on a site and $\textbf{d}$ is the displacement vector to a neighboring site. For the upper branch experiments, $t_n/t \sim -8\%$ and for the lower branch experiments $t_n/t \sim -16\%$.  To study the effect of the density-dependent tunneling on the measured CDW correlations and singles fraction, we performed Numerical Linked Cluster Expansion (NLCE) simulations where the tunneling was replaced by $t_{\sigma,\textbf{r},\textbf{r}'} = t + (n_{-\sigma,\textbf{r}}+n_{-\sigma,\textbf{r}'}) t_n$, where if $\sigma$=$\uparrow$, $-\sigma$=$\downarrow$ and vice versa. In other words, the tunneling of a spin $\sigma$ from site $\textbf{r}$ to $\textbf{r}'$ depends on the total density of the opposite spin at the two sites between which the tunneling process is taking place. The strongest effect was about $10\%$ on the singles fraction and that would change the fit value of the data in Fig.~4b,c to $U/t=-5.4$ and $T=0.4t$. Both this value of $U/t$ and the one obtained without corrections to the Hubbard model fall within our experimental uncertainties.}
\end{itemize}

{\bf Determinantal Quantum Monte Carlo calculations} The DQMC simulations shown in the paper were all performed using a Fortran 90/95 package called QUantum Electron Simulation Toolbox (QUEST) developed and maintained by R. T. Scalettar et. al. \cite{Varney2009}. For a spin-balanced system with attractive interactions ($U/t < 0$) the calculations do not suffer from a fermion sign problem.

Simulations for Fig.~2 and Fig.~3 of the main text were performed on a square lattice of 8$\times$8 sites with $U/t = -5.7$ and a chemical potential ($\mu /t$) varying from -3 to 1.5 with $\mu=0$ representing half filling. The inverse temperature $\beta = Ld\tau$  was split into $L=40$ imaginary time slices, with an interval $d\tau = 0.0556$. To obtain higher statistics, the simulations were run over 100,000 passes. For Fig.~4 of the main text, $U/t$ was fixed at -5.7 and $\mu$ to zero. The temperature was scanned by varying $d\tau$ from 0.0046 to 0.083 for a fixed $L=40$ and each point was averaged over 100,000 passes.

The DQMC simulations in the paper are performed using homogeneous systems. We rely on the local density approximation (LDA) for comparison to the experiment where the density varies slowly due to the harmonic trapping potential. For data shown in Fig.~\ref{fig:corr_U}, the density gradient at $n=1$ varied from 0.07 atoms/(site)$^2$ for a depth of $4.1~E_R$ to 0.14 atoms/(site)$^2$ for a depth of $7.4~E_R$. To verify that the LDA holds for our system, we have performed DQMC simulations in the presence of a linearly varying chemical potential along one direction of the 2D Hubbard lattice to reproduce the maximal density gradients observed in the experiment. The results of this calculation showed that the LDA holds to excellent approximation for our experimental parameters. However, we note that violation of the LDA had been predicted for such a system at much lower temperatures and higher gradients \cite{Assmann2012}.


\begin{thebibliography}{46}%
\makeatletter
\providecommand \@ifxundefined [1]{%
 \@ifx{#1\undefined}
}%
\providecommand \@ifnum [1]{%
 \ifnum #1\expandafter \@firstoftwo
 \else \expandafter \@secondoftwo
 \fi
}%
\providecommand \@ifx [1]{%
 \ifx #1\expandafter \@firstoftwo
 \else \expandafter \@secondoftwo
 \fi
}%
\providecommand \natexlab [1]{#1}%
\providecommand \enquote  [1]{``#1''}%
\providecommand \bibnamefont  [1]{#1}%
\providecommand \bibfnamefont [1]{#1}%
\providecommand \citenamefont [1]{#1}%
\providecommand \href@noop [0]{\@secondoftwo}%
\providecommand \href [0]{\begingroup \@sanitize@url \@href}%
\providecommand \@href[1]{\@@startlink{#1}\@@href}%
\providecommand \@@href[1]{\endgroup#1\@@endlink}%
\providecommand \@sanitize@url [0]{\catcode `\\12\catcode `\$12\catcode
  `\&12\catcode `\#12\catcode `\^12\catcode `\_12\catcode `\%12\relax}%
\providecommand \@@startlink[1]{}%
\providecommand \@@endlink[0]{}%
\providecommand \url  [0]{\begingroup\@sanitize@url \@url }%
\providecommand \@url [1]{\endgroup\@href {#1}{\urlprefix }}%
\providecommand \urlprefix  [0]{URL }%
\providecommand \Eprint [0]{\href }%
\providecommand \doibase [0]{http://dx.doi.org/}%
\providecommand \selectlanguage [0]{\@gobble}%
\providecommand \bibinfo  [0]{\@secondoftwo}%
\providecommand \bibfield  [0]{\@secondoftwo}%
\providecommand \translation [1]{[#1]}%
\providecommand \BibitemOpen [0]{}%
\providecommand \bibitemStop [0]{}%
\providecommand \bibitemNoStop [0]{.\EOS\space}%
\providecommand \EOS [0]{\spacefactor3000\relax}%
\providecommand \BibitemShut  [1]{\csname bibitem#1\endcsname}%
\let\auto@bib@innerbib\@empty
\bibitem [{\citenamefont {Micnas}\ \emph {et~al.}(1990)\citenamefont {Micnas},
  \citenamefont {Ranninger},\ and\ \citenamefont {Robaszkiewicz}}]{Micnas1990}%
  \BibitemOpen
  \bibfield  {author} {\bibinfo {author} {\bibfnamefont {R.}~\bibnamefont
  {Micnas}}, \bibinfo {author} {\bibfnamefont {J.}~\bibnamefont {Ranninger}}, \
  and\ \bibinfo {author} {\bibfnamefont {S.}~\bibnamefont {Robaszkiewicz}},\
  }\href {\doibase 10.1103/RevModPhys.62.113} {\bibfield  {journal} {\bibinfo
  {journal} {Rev. Mod. Phys.}\ }\textbf {\bibinfo {volume} {62}},\ \bibinfo
  {pages} {113} (\bibinfo {year} {1990})}\BibitemShut {NoStop}%
\bibitem [{\citenamefont {Randeria}\ \emph {et~al.}(1992)\citenamefont
  {Randeria}, \citenamefont {Trivedi}, \citenamefont {Moreo},\ and\
  \citenamefont {Scalettar}}]{Randeria1992}%
  \BibitemOpen
  \bibfield  {author} {\bibinfo {author} {\bibfnamefont {M.}~\bibnamefont
  {Randeria}}, \bibinfo {author} {\bibfnamefont {N.}~\bibnamefont {Trivedi}},
  \bibinfo {author} {\bibfnamefont {A.}~\bibnamefont {Moreo}}, \ and\ \bibinfo
  {author} {\bibfnamefont {R.~T.}\ \bibnamefont {Scalettar}},\ }\href {\doibase
  10.1103/PhysRevLett.69.2001} {\bibfield  {journal} {\bibinfo  {journal}
  {Phys. Rev. Lett.}\ }\textbf {\bibinfo {volume} {69}},\ \bibinfo {pages}
  {2001} (\bibinfo {year} {1992})}\BibitemShut {NoStop}%
\bibitem [{\citenamefont {Trivedi}\ and\ \citenamefont
  {Randeria}(1995)}]{Trivedi1995}%
  \BibitemOpen
  \bibfield  {author} {\bibinfo {author} {\bibfnamefont {N.}~\bibnamefont
  {Trivedi}}\ and\ \bibinfo {author} {\bibfnamefont {M.}~\bibnamefont
  {Randeria}},\ }\href {\doibase 10.1103/PhysRevLett.75.312} {\bibfield
  {journal} {\bibinfo  {journal} {Phys. Rev. Lett.}\ }\textbf {\bibinfo
  {volume} {75}},\ \bibinfo {pages} {312} (\bibinfo {year} {1995})}\BibitemShut
  {NoStop}%
\bibitem [{\citenamefont {Singer}\ \emph {et~al.}(1996)\citenamefont {Singer},
  \citenamefont {Pedersen}, \citenamefont {Schneider}, \citenamefont {Beck},\
  and\ \citenamefont {Matuttis}}]{Singer1996}%
  \BibitemOpen
  \bibfield  {author} {\bibinfo {author} {\bibfnamefont {J.~M.}\ \bibnamefont
  {Singer}}, \bibinfo {author} {\bibfnamefont {M.~H.}\ \bibnamefont
  {Pedersen}}, \bibinfo {author} {\bibfnamefont {T.}~\bibnamefont {Schneider}},
  \bibinfo {author} {\bibfnamefont {H.}~\bibnamefont {Beck}}, \ and\ \bibinfo
  {author} {\bibfnamefont {H.-G.}\ \bibnamefont {Matuttis}},\ }\href {\doibase
  10.1103/PhysRevB.54.1286} {\bibfield  {journal} {\bibinfo  {journal} {Phys.
  Rev. B}\ }\textbf {\bibinfo {volume} {54}},\ \bibinfo {pages} {1286}
  (\bibinfo {year} {1996})}\BibitemShut {NoStop}%
\bibitem [{\citenamefont {Kyung}\ \emph {et~al.}(2001)\citenamefont {Kyung},
  \citenamefont {Allen},\ and\ \citenamefont {Tremblay}}]{Kyung2001}%
  \BibitemOpen
  \bibfield  {author} {\bibinfo {author} {\bibfnamefont {B.}~\bibnamefont
  {Kyung}}, \bibinfo {author} {\bibfnamefont {S.}~\bibnamefont {Allen}}, \ and\
  \bibinfo {author} {\bibfnamefont {A.-M.~S.}\ \bibnamefont {Tremblay}},\
  }\href {\doibase 10.1103/PhysRevB.64.075116} {\bibfield  {journal} {\bibinfo
  {journal} {Phys. Rev. B}\ }\textbf {\bibinfo {volume} {64}},\ \bibinfo
  {pages} {075116} (\bibinfo {year} {2001})}\BibitemShut {NoStop}%
\bibitem [{\citenamefont {Yang}\ and\ \citenamefont {Zhang}(1990)}]{Yang1990}%
  \BibitemOpen
  \bibfield  {author} {\bibinfo {author} {\bibfnamefont {C. N.}~\bibnamefont
  {Yang}}\ and\ \bibinfo {author} {\bibfnamefont {S. C.}~\bibnamefont {Zhang}},\
  }\href {\doibase 10.1142/S0217984990000933} {\bibfield  {journal} {\bibinfo
  {journal} {Mod. Phys. Lett. B}\ }\textbf {\bibinfo {volume} {04}},\ \bibinfo
  {pages} {759} (\bibinfo {year} {1990})}\BibitemShut {NoStop}%
\bibitem [{\citenamefont {Hubbard}(1963)}]{Hubbard1963}%
  \BibitemOpen
  \bibfield  {author} {\bibinfo {author} {\bibfnamefont {J.}~\bibnamefont
  {Hubbard}},\ }\href {\doibase 10.1098/rspa.1963.0204} {\bibfield  {journal}
  {\bibinfo  {journal} {Proc. Roy. Soc. A}\ }\textbf {\bibinfo {volume}
  {276}},\ \bibinfo {pages} {238} (\bibinfo {year} {1963})}\BibitemShut
  {NoStop}%
\bibitem [{\citenamefont {Auerbach}()}]{Auerbach1990}%
  \BibitemOpen
  \bibfield  {author} {\bibinfo {author} {\bibfnamefont {A.}~\bibnamefont
  {Auerbach}},\ }\href {\doibase 10.1007/978-1-4612-0869-3} {\emph {\bibinfo
  {title} {Interacting Electrons and Quantum Magnetism}}}\ (\bibinfo
  {publisher} {Springer, New York, 1994})\BibitemShut {NoStop}%
\bibitem [{\citenamefont {J\"ordens}\ \emph {et~al.}(2008)\citenamefont
  {J\"ordens}, \citenamefont {Strohmaier}, \citenamefont {G\"unter},
  \citenamefont {Moritz},\ and\ \citenamefont {Esslinger}}]{Joerdens2008}%
  \BibitemOpen
  \bibfield  {author} {\bibinfo {author} {\bibfnamefont {R.}~\bibnamefont
  {J\"ordens}}, \bibinfo {author} {\bibfnamefont {N.}~\bibnamefont
  {Strohmaier}}, \bibinfo {author} {\bibfnamefont {K.}~\bibnamefont
  {G\"unter}}, \bibinfo {author} {\bibfnamefont {H.}~\bibnamefont {Moritz}}, \
  and\ \bibinfo {author} {\bibfnamefont {T.}~\bibnamefont {Esslinger}},\ }\href
  {\doibase 10.1038/nature07244} {\bibfield  {journal} {\bibinfo  {journal}
  {Nature}\ }\textbf {\bibinfo {volume} {455}},\ \bibinfo {pages} {204}
  (\bibinfo {year} {2008})}\BibitemShut {NoStop}%
\bibitem [{\citenamefont {Schneider}\ \emph {et~al.}(2008)\citenamefont
  {Schneider}, \citenamefont {Hackerm\"uller}, \citenamefont {Will},
  \citenamefont {Best}, \citenamefont {Bloch}, \citenamefont {Costi},
  \citenamefont {Helmes}, \citenamefont {Rasch},\ and\ \citenamefont
  {Rosch}}]{Schneider2008}%
  \BibitemOpen
  \bibfield  {author} {\bibinfo {author} {\bibfnamefont {U.}~\bibnamefont
  {Schneider}}, \bibinfo {author} {\bibfnamefont {L.}~\bibnamefont
  {Hackerm\"uller}}, \bibinfo {author} {\bibfnamefont {S.}~\bibnamefont
  {Will}}, \bibinfo {author} {\bibfnamefont {T.}~\bibnamefont {Best}}, \bibinfo
  {author} {\bibfnamefont {I.}~\bibnamefont {Bloch}}, \bibinfo {author}
  {\bibfnamefont {T.~A.}\ \bibnamefont {Costi}}, \bibinfo {author}
  {\bibfnamefont {R.~W.}\ \bibnamefont {Helmes}}, \bibinfo {author}
  {\bibfnamefont {D.}~\bibnamefont {Rasch}}, \ and\ \bibinfo {author}
  {\bibfnamefont {A.}~\bibnamefont {Rosch}},\ }\href {\doibase
  10.1126/science.1165449} {\bibfield  {journal} {\bibinfo  {journal}
  {Science}\ }\textbf {\bibinfo {volume} {322}},\ \bibinfo {pages} {1520}
  (\bibinfo {year} {2008})}\BibitemShut {NoStop}%
\bibitem [{\citenamefont {Greif}\ \emph {et~al.}(2013)\citenamefont {Greif},
  \citenamefont {Uehlinger}, \citenamefont {Jotzu}, \citenamefont {Tarruell},\
  and\ \citenamefont {Esslinger}}]{Greif2013}%
  \BibitemOpen
  \bibfield  {author} {\bibinfo {author} {\bibfnamefont {D.}~\bibnamefont
  {Greif}}, \bibinfo {author} {\bibfnamefont {T.}~\bibnamefont {Uehlinger}},
  \bibinfo {author} {\bibfnamefont {G.}~\bibnamefont {Jotzu}}, \bibinfo
  {author} {\bibfnamefont {L.}~\bibnamefont {Tarruell}}, \ and\ \bibinfo
  {author} {\bibfnamefont {T.}~\bibnamefont {Esslinger}},\ }\href {\doibase
  10.1126/science.1236362} {\bibfield  {journal} {\bibinfo  {journal}
  {Science}\ }\textbf {\bibinfo {volume} {340}},\ \bibinfo {pages} {1307}
  (\bibinfo {year} {2013})}\BibitemShut {NoStop}%
\bibitem [{\citenamefont {Hart}\ \emph {et~al.}(2015)\citenamefont {Hart},
  \citenamefont {Duarte}, \citenamefont {Yang}, \citenamefont {Liu},
  \citenamefont {Paiva}, \citenamefont {Khatami}, \citenamefont {Scalettar},
  \citenamefont {Trivedi}, \citenamefont {Huse},\ and\ \citenamefont
  {Hulet}}]{Hart2015}%
  \BibitemOpen
  \bibfield  {author} {\bibinfo {author} {\bibfnamefont {R.~A.}\ \bibnamefont
  {Hart}}, \bibinfo {author} {\bibfnamefont {P.~M.}\ \bibnamefont {Duarte}},
  \bibinfo {author} {\bibfnamefont {T.-L.}\ \bibnamefont {Yang}}, \bibinfo
  {author} {\bibfnamefont {X.}~\bibnamefont {Liu}}, \bibinfo {author}
  {\bibfnamefont {T.}~\bibnamefont {Paiva}}, \bibinfo {author} {\bibfnamefont
  {E.}~\bibnamefont {Khatami}}, \bibinfo {author} {\bibfnamefont {R.~T.}\
  \bibnamefont {Scalettar}}, \bibinfo {author} {\bibfnamefont {N.}~\bibnamefont
  {Trivedi}}, \bibinfo {author} {\bibfnamefont {D.~A.}\ \bibnamefont {Huse}}, \
  and\ \bibinfo {author} {\bibfnamefont {R.~G.}\ \bibnamefont {Hulet}},\ }\href
  {http://dx.doi.org/10.1038/nature14223} {\bibfield  {journal} {\bibinfo
  {journal} {Nature}\ }\textbf {\bibinfo {volume} {519}},\ \bibinfo {pages}
  {211} (\bibinfo {year} {2015})}\BibitemShut {NoStop}%
\bibitem [{\citenamefont {Parsons}\ \emph {et~al.}(2016)\citenamefont
  {Parsons}, \citenamefont {Mazurenko}, \citenamefont {Chiu}, \citenamefont
  {Ji}, \citenamefont {Greif},\ and\ \citenamefont {Greiner}}]{Parsons2016}%
  \BibitemOpen
  \bibfield  {author} {\bibinfo {author} {\bibfnamefont {M.~F.}\ \bibnamefont
  {Parsons}}, \bibinfo {author} {\bibfnamefont {A.}~\bibnamefont {Mazurenko}},
  \bibinfo {author} {\bibfnamefont {C.~S.}\ \bibnamefont {Chiu}}, \bibinfo
  {author} {\bibfnamefont {G.}~\bibnamefont {Ji}}, \bibinfo {author}
  {\bibfnamefont {D.}~\bibnamefont {Greif}}, \ and\ \bibinfo {author}
  {\bibfnamefont {M.}~\bibnamefont {Greiner}},\ }\href {\doibase
  10.1126/science.aag1430} {\bibfield  {journal} {\bibinfo  {journal}
  {Science}\ }\textbf {\bibinfo {volume} {353}},\ \bibinfo {pages} {1253}
  (\bibinfo {year} {2016})}\BibitemShut {NoStop}%
\bibitem [{\citenamefont {Cheuk}\ \emph {et~al.}(2016)\citenamefont {Cheuk},
  \citenamefont {Nichols}, \citenamefont {Lawrence}, \citenamefont {Okan},
  \citenamefont {Zhang}, \citenamefont {Khatami}, \citenamefont {Trivedi},
  \citenamefont {Paiva}, \citenamefont {Rigol},\ and\ \citenamefont
  {Zwierlein}}]{Cheuk2016}%
  \BibitemOpen
  \bibfield  {author} {\bibinfo {author} {\bibfnamefont {L.~W.}\ \bibnamefont
  {Cheuk}}, \bibinfo {author} {\bibfnamefont {M.~A.}\ \bibnamefont {Nichols}},
  \bibinfo {author} {\bibfnamefont {K.~R.}\ \bibnamefont {Lawrence}}, \bibinfo
  {author} {\bibfnamefont {M.}~\bibnamefont {Okan}}, \bibinfo {author}
  {\bibfnamefont {H.}~\bibnamefont {Zhang}}, \bibinfo {author} {\bibfnamefont
  {E.}~\bibnamefont {Khatami}}, \bibinfo {author} {\bibfnamefont
  {N.}~\bibnamefont {Trivedi}}, \bibinfo {author} {\bibfnamefont
  {T.}~\bibnamefont {Paiva}}, \bibinfo {author} {\bibfnamefont
  {M.}~\bibnamefont {Rigol}}, \ and\ \bibinfo {author} {\bibfnamefont {M.~W.}\
  \bibnamefont {Zwierlein}},\ }\href {\doibase 10.1126/science.aag3349}
  {\bibfield  {journal} {\bibinfo  {journal} {Science}\ }\textbf {\bibinfo
  {volume} {353}},\ \bibinfo {pages} {1260} (\bibinfo {year}
  {2016})}\BibitemShut {NoStop}%
\bibitem [{\citenamefont {Boll}\ \emph {et~al.}(2016)\citenamefont {Boll},
  \citenamefont {Hilker}, \citenamefont {Salomon}, \citenamefont {Omran},
  \citenamefont {Nespolo}, \citenamefont {Pollet}, \citenamefont {Bloch},\ and\
  \citenamefont {Gross}}]{Boll2016}%
  \BibitemOpen
  \bibfield  {author} {\bibinfo {author} {\bibfnamefont {M.}~\bibnamefont
  {Boll}}, \bibinfo {author} {\bibfnamefont {T.~A.}\ \bibnamefont {Hilker}},
  \bibinfo {author} {\bibfnamefont {G.}~\bibnamefont {Salomon}}, \bibinfo
  {author} {\bibfnamefont {A.}~\bibnamefont {Omran}}, \bibinfo {author}
  {\bibfnamefont {J.}~\bibnamefont {Nespolo}}, \bibinfo {author} {\bibfnamefont
  {L.}~\bibnamefont {Pollet}}, \bibinfo {author} {\bibfnamefont
  {I.}~\bibnamefont {Bloch}}, \ and\ \bibinfo {author} {\bibfnamefont
  {C.}~\bibnamefont {Gross}},\ }\href {\doibase 10.1126/science.aag1635}
  {\bibfield  {journal} {\bibinfo  {journal} {Science}\ }\textbf {\bibinfo
  {volume} {353}},\ \bibinfo {pages} {1257} (\bibinfo {year}
  {2016})}\BibitemShut {NoStop}%
\bibitem [{\citenamefont {Brown}\ \emph {et~al.}(2016)\citenamefont {Brown},
  \citenamefont {Mitra}, \citenamefont {Guardado-Sanchez}, \citenamefont
  {Schau\ss}, \citenamefont {Kondov}, \citenamefont {Khatami}, \citenamefont
  {Paiva}, \citenamefont {Trivedi}, \citenamefont {Huse},\ and\ \citenamefont
  {Bakr}}]{Brown2016}%
  \BibitemOpen
  \bibfield  {author} {\bibinfo {author} {\bibfnamefont {P.~T.}\ \bibnamefont
  {Brown}}, \bibinfo {author} {\bibfnamefont {D.}~\bibnamefont {Mitra}},
  \bibinfo {author} {\bibfnamefont {E.}~\bibnamefont {Guardado-Sanchez}},
  \bibinfo {author} {\bibfnamefont {P.}~\bibnamefont {Schau\ss}}, \bibinfo
  {author} {\bibfnamefont {S.~S.}\ \bibnamefont {Kondov}}, \bibinfo {author}
  {\bibfnamefont {E.}~\bibnamefont {Khatami}}, \bibinfo {author} {\bibfnamefont
  {T.}~\bibnamefont {Paiva}}, \bibinfo {author} {\bibfnamefont
  {N.}~\bibnamefont {Trivedi}}, \bibinfo {author} {\bibfnamefont {D.~A.}\
  \bibnamefont {Huse}}, \ and\ \bibinfo {author} {\bibfnamefont {W.~S.}\
  \bibnamefont {Bakr}},\ }\href {http://arxiv.org/abs/1612.07746} {\bibfield
  {journal} {\bibinfo  {journal} {arXiv:1612.07746}\ } (\bibinfo {year}
  {2016})}\BibitemShut {NoStop}%
\bibitem [{\citenamefont {Cocchi}\ \emph {et~al.}(2016)\citenamefont {Cocchi},
  \citenamefont {Miller}, \citenamefont {Drewes}, \citenamefont {Koschorreck},
  \citenamefont {Pertot}, \citenamefont {Brennecke},\ and\ \citenamefont
  {K\"ohl}}]{Cocchi2016}%
  \BibitemOpen
  \bibfield  {author} {\bibinfo {author} {\bibfnamefont {E.}~\bibnamefont
  {Cocchi}}, \bibinfo {author} {\bibfnamefont {L.~A.}\ \bibnamefont {Miller}},
  \bibinfo {author} {\bibfnamefont {J.~H.}\ \bibnamefont {Drewes}}, \bibinfo
  {author} {\bibfnamefont {M.}~\bibnamefont {Koschorreck}}, \bibinfo {author}
  {\bibfnamefont {D.}~\bibnamefont {Pertot}}, \bibinfo {author} {\bibfnamefont
  {F.}~\bibnamefont {Brennecke}}, \ and\ \bibinfo {author} {\bibfnamefont
  {M.}~\bibnamefont {K\"ohl}},\ }\href {\doibase
  10.1103/PhysRevLett.116.175301} {\bibfield  {journal} {\bibinfo  {journal}
  {Phys. Rev. Lett.}\ }\textbf {\bibinfo {volume} {116}},\ \bibinfo {pages}
  {175301} (\bibinfo {year} {2016})}\BibitemShut {NoStop}%
\bibitem [{\citenamefont {Strohmaier}\ \emph {et~al.}(2007)\citenamefont
  {Strohmaier}, \citenamefont {Takasu}, \citenamefont {G\"unter}, \citenamefont
  {J\"ordens}, \citenamefont {K\"ohl}, \citenamefont {Moritz},\ and\
  \citenamefont {Esslinger}}]{Strohmaier2007}%
  \BibitemOpen
  \bibfield  {author} {\bibinfo {author} {\bibfnamefont {N.}~\bibnamefont
  {Strohmaier}}, \bibinfo {author} {\bibfnamefont {Y.}~\bibnamefont {Takasu}},
  \bibinfo {author} {\bibfnamefont {K.}~\bibnamefont {G\"unter}}, \bibinfo
  {author} {\bibfnamefont {R.}~\bibnamefont {J\"ordens}}, \bibinfo {author}
  {\bibfnamefont {M.}~\bibnamefont {K\"ohl}}, \bibinfo {author} {\bibfnamefont
  {H.}~\bibnamefont {Moritz}}, \ and\ \bibinfo {author} {\bibfnamefont
  {T.}~\bibnamefont {Esslinger}},\ }\href {\doibase
  10.1103/PhysRevLett.99.220601} {\bibfield  {journal} {\bibinfo  {journal}
  {Phys. Rev. Lett.}\ }\textbf {\bibinfo {volume} {99}},\ \bibinfo {pages}
  {220601} (\bibinfo {year} {2007})}\BibitemShut {NoStop}%
\bibitem [{\citenamefont {Hackerm\"uller}\ \emph {et~al.}(2010)\citenamefont
  {Hackerm\"uller}, \citenamefont {Schneider}, \citenamefont {Moreno-Cardoner},
  \citenamefont {Kitagawa}, \citenamefont {Best}, \citenamefont {Will},
  \citenamefont {Demler}, \citenamefont {Altman}, \citenamefont {Bloch},\ and\
  \citenamefont {Paredes}}]{Hackermueller2010}%
  \BibitemOpen
  \bibfield  {author} {\bibinfo {author} {\bibfnamefont {L.}~\bibnamefont
  {Hackerm\"uller}}, \bibinfo {author} {\bibfnamefont {U.}~\bibnamefont
  {Schneider}}, \bibinfo {author} {\bibfnamefont {M.}~\bibnamefont
  {Moreno-Cardoner}}, \bibinfo {author} {\bibfnamefont {T.}~\bibnamefont
  {Kitagawa}}, \bibinfo {author} {\bibfnamefont {T.}~\bibnamefont {Best}},
  \bibinfo {author} {\bibfnamefont {S.}~\bibnamefont {Will}}, \bibinfo {author}
  {\bibfnamefont {E.}~\bibnamefont {Demler}}, \bibinfo {author} {\bibfnamefont
  {E.}~\bibnamefont {Altman}}, \bibinfo {author} {\bibfnamefont
  {I.}~\bibnamefont {Bloch}}, \ and\ \bibinfo {author} {\bibfnamefont
  {B.}~\bibnamefont {Paredes}},\ }\href {\doibase 10.1126/science.1184565}
  {\bibfield  {journal} {\bibinfo  {journal} {Science}\ }\textbf {\bibinfo
  {volume} {327}},\ \bibinfo {pages} {1621} (\bibinfo {year}
  {2010})}\BibitemShut {NoStop}%
\bibitem [{\citenamefont {Schneider}\ \emph {et~al.}(2012)\citenamefont
  {Schneider}, \citenamefont {Hackerm\"uller}, \citenamefont {Ronzheimer},
  \citenamefont {Will}, \citenamefont {Braun}, \citenamefont {Best},
  \citenamefont {Bloch}, \citenamefont {Demler}, \citenamefont {Mandt},
  \citenamefont {Rasch},\ and\ \citenamefont {Rosch}}]{Schneider2012}%
  \BibitemOpen
  \bibfield  {author} {\bibinfo {author} {\bibfnamefont {U.}~\bibnamefont
  {Schneider}}, \bibinfo {author} {\bibfnamefont {L.}~\bibnamefont
  {Hackerm\"uller}}, \bibinfo {author} {\bibfnamefont {J.~P.}\ \bibnamefont
  {Ronzheimer}}, \bibinfo {author} {\bibfnamefont {S.}~\bibnamefont {Will}},
  \bibinfo {author} {\bibfnamefont {S.}~\bibnamefont {Braun}}, \bibinfo
  {author} {\bibfnamefont {T.}~\bibnamefont {Best}}, \bibinfo {author}
  {\bibfnamefont {I.}~\bibnamefont {Bloch}}, \bibinfo {author} {\bibfnamefont
  {E.}~\bibnamefont {Demler}}, \bibinfo {author} {\bibfnamefont
  {S.}~\bibnamefont {Mandt}}, \bibinfo {author} {\bibfnamefont
  {D.}~\bibnamefont {Rasch}}, \ and\ \bibinfo {author} {\bibfnamefont
  {A.}~\bibnamefont {Rosch}},\ }\href {\doibase 10.1038/nphys2205} {\bibfield
  {journal} {\bibinfo  {journal} {Nat Phys}\ }\textbf {\bibinfo {volume} {8}},\
  \bibinfo {pages} {213} (\bibinfo {year} {2012})}\BibitemShut {NoStop}%
\bibitem [{Ing(2008)}]{Inguscio2008}%
  \BibitemOpen
  \href@noop {} {\emph {\bibinfo {title} {Proceedings of the International
  School of Physics ``Enrico Fermi", Course CLXIV}}}\ (\bibinfo  {publisher}
  {IOS Press, Amsterdam},\ \bibinfo {year} {2008})\BibitemShut {NoStop}%
\bibitem [{\citenamefont {Chin}\ \emph {et~al.}(2006)\citenamefont {Chin},
  \citenamefont {Miller}, \citenamefont {Liu}, \citenamefont {Stan},
  \citenamefont {Setiawan}, \citenamefont {Sanner}, \citenamefont {Xu},\ and\
  \citenamefont {Ketterle}}]{Chin2006}%
  \BibitemOpen
  \bibfield  {author} {\bibinfo {author} {\bibfnamefont {J.~K.}\ \bibnamefont
  {Chin}}, \bibinfo {author} {\bibfnamefont {D.~E.}\ \bibnamefont {Miller}},
  \bibinfo {author} {\bibfnamefont {Y.}~\bibnamefont {Liu}}, \bibinfo {author}
  {\bibfnamefont {C.}~\bibnamefont {Stan}}, \bibinfo {author} {\bibfnamefont
  {W.}~\bibnamefont {Setiawan}}, \bibinfo {author} {\bibfnamefont
  {C.}~\bibnamefont {Sanner}}, \bibinfo {author} {\bibfnamefont
  {K.}~\bibnamefont {Xu}}, \ and\ \bibinfo {author} {\bibfnamefont
  {W.}~\bibnamefont {Ketterle}},\ }\href {\doibase 10.1038/nature05224}
  {\bibfield  {journal} {\bibinfo  {journal} {Nature}\ }\textbf {\bibinfo
  {volume} {443}},\ \bibinfo {pages} {961} (\bibinfo {year}
  {2006})}\BibitemShut {NoStop}%
\bibitem [{\citenamefont {Duan}(2005)}]{Duan2005}%
  \BibitemOpen
  \bibfield  {author} {\bibinfo {author} {\bibfnamefont {L.-M.}\ \bibnamefont
  {Duan}},\ }\href {\doibase 10.1103/PhysRevLett.95.243202} {\bibfield
  {journal} {\bibinfo  {journal} {Phys. Rev. Lett.}\ }\textbf {\bibinfo
  {volume} {95}},\ \bibinfo {pages} {243202} (\bibinfo {year}
  {2005})}\BibitemShut {NoStop}%
\bibitem [{\citenamefont {Carr}\ and\ \citenamefont
  {Holland}(2005)}]{Carr2005}%
  \BibitemOpen
  \bibfield  {author} {\bibinfo {author} {\bibfnamefont {L.~D.}\ \bibnamefont
  {Carr}}\ and\ \bibinfo {author} {\bibfnamefont {M.~J.}\ \bibnamefont
  {Holland}},\ }\href {\doibase 10.1103/PhysRevA.72.031604} {\bibfield
  {journal} {\bibinfo  {journal} {Phys. Rev. A}\ }\textbf {\bibinfo {volume}
  {72}},\ \bibinfo {pages} {031604} (\bibinfo {year} {2005})}\BibitemShut
  {NoStop}%
\bibitem [{\citenamefont {Zhou}(2005)}]{Zhou2005}%
  \BibitemOpen
  \bibfield  {author} {\bibinfo {author} {\bibfnamefont {F.}~\bibnamefont
  {Zhou}},\ }\href {\doibase 10.1103/PhysRevB.72.220501} {\bibfield  {journal}
  {\bibinfo  {journal} {Phys. Rev. B}\ }\textbf {\bibinfo {volume} {72}},\
  \bibinfo {pages} {220501} (\bibinfo {year} {2005})}\BibitemShut {NoStop}%
\bibitem [{\citenamefont {Diener}\ and\ \citenamefont {Ho}(2006)}]{Diener2006}%
  \BibitemOpen
  \bibfield  {author} {\bibinfo {author} {\bibfnamefont {R.~B.}\ \bibnamefont
  {Diener}}\ and\ \bibinfo {author} {\bibfnamefont {T.-L.}\ \bibnamefont
  {Ho}},\ }\href {\doibase 10.1103/PhysRevLett.96.010402} {\bibfield  {journal}
  {\bibinfo  {journal} {Phys. Rev. Lett.}\ }\textbf {\bibinfo {volume} {96}},\
  \bibinfo {pages} {010402} (\bibinfo {year} {2006})}\BibitemShut {NoStop}%
\bibitem [{\citenamefont {Hirsch}(1985)}]{Hirsch1985}%
  \BibitemOpen
  \bibfield  {author} {\bibinfo {author} {\bibfnamefont {J.~E.}\ \bibnamefont
  {Hirsch}},\ }\href {\doibase 10.1103/PhysRevB.31.4403} {\bibfield  {journal}
  {\bibinfo  {journal} {Phys. Rev. B}\ }\textbf {\bibinfo {volume} {31}},\
  \bibinfo {pages} {4403} (\bibinfo {year} {1985})}\BibitemShut {NoStop}%
\bibitem [{\citenamefont {Scalettar}\ \emph {et~al.}(1989)\citenamefont
  {Scalettar}, \citenamefont {Loh}, \citenamefont {Gubernatis}, \citenamefont
  {Moreo}, \citenamefont {White}, \citenamefont {Scalapino}, \citenamefont
  {Sugar},\ and\ \citenamefont {Dagotto}}]{Scalettar1989}%
  \BibitemOpen
  \bibfield  {author} {\bibinfo {author} {\bibfnamefont {R.~T.}\ \bibnamefont
  {Scalettar}}, \bibinfo {author} {\bibfnamefont {E.~Y.}\ \bibnamefont {Loh}},
  \bibinfo {author} {\bibfnamefont {J.~E.}\ \bibnamefont {Gubernatis}},
  \bibinfo {author} {\bibfnamefont {A.}~\bibnamefont {Moreo}}, \bibinfo
  {author} {\bibfnamefont {S.~R.}\ \bibnamefont {White}}, \bibinfo {author}
  {\bibfnamefont {D.~J.}\ \bibnamefont {Scalapino}}, \bibinfo {author}
  {\bibfnamefont {R.~L.}\ \bibnamefont {Sugar}}, \ and\ \bibinfo {author}
  {\bibfnamefont {E.}~\bibnamefont {Dagotto}},\ }\href {\doibase
  10.1103/PhysRevLett.62.1407} {\bibfield  {journal} {\bibinfo  {journal}
  {Phys. Rev. Lett.}\ }\textbf {\bibinfo {volume} {62}},\ \bibinfo {pages}
  {1407} (\bibinfo {year} {1989})}\BibitemShut {NoStop}%
\bibitem [{\citenamefont {Moreo}\ and\ \citenamefont
  {Scalapino}(1991)}]{Moreo1991}%
  \BibitemOpen
  \bibfield  {author} {\bibinfo {author} {\bibfnamefont {A.}~\bibnamefont
  {Moreo}}\ and\ \bibinfo {author} {\bibfnamefont {D.~J.}\ \bibnamefont
  {Scalapino}},\ }\href {\doibase 10.1103/PhysRevLett.66.946} {\bibfield
  {journal} {\bibinfo  {journal} {Phys. Rev. Lett.}\ }\textbf {\bibinfo
  {volume} {66}},\ \bibinfo {pages} {946} (\bibinfo {year} {1991})}\BibitemShut
  {NoStop}%
\bibitem [{\citenamefont {Paiva}\ \emph {et~al.}(2004)\citenamefont {Paiva},
  \citenamefont {dos Santos}, \citenamefont {Scalettar},\ and\ \citenamefont
  {Denteneer}}]{Paiva2004}%
  \BibitemOpen
  \bibfield  {author} {\bibinfo {author} {\bibfnamefont {T.}~\bibnamefont
  {Paiva}}, \bibinfo {author} {\bibfnamefont {R.~R.}\ \bibnamefont {dos
  Santos}}, \bibinfo {author} {\bibfnamefont {R.~T.}\ \bibnamefont
  {Scalettar}}, \ and\ \bibinfo {author} {\bibfnamefont {P.~J.~H.}\
  \bibnamefont {Denteneer}},\ }\href {\doibase 10.1103/PhysRevB.69.184501}
  {\bibfield  {journal} {\bibinfo  {journal} {Phys. Rev. B}\ }\textbf {\bibinfo
  {volume} {69}},\ \bibinfo {pages} {184501} (\bibinfo {year}
  {2004})}\BibitemShut {NoStop}%
\bibitem [{\citenamefont {Haller}\ \emph {et~al.}(2015)\citenamefont {Haller},
  \citenamefont {Hudson}, \citenamefont {Kelly}, \citenamefont {Cotta},
  \citenamefont {Peaudecerf}, \citenamefont {Bruce},\ and\ \citenamefont
  {Kuhr}}]{Haller2015}%
  \BibitemOpen
  \bibfield  {author} {\bibinfo {author} {\bibfnamefont {E.}~\bibnamefont
  {Haller}}, \bibinfo {author} {\bibfnamefont {J.}~\bibnamefont {Hudson}},
  \bibinfo {author} {\bibfnamefont {A.}~\bibnamefont {Kelly}}, \bibinfo
  {author} {\bibfnamefont {D.~A.}\ \bibnamefont {Cotta}}, \bibinfo {author}
  {\bibfnamefont {B.}~\bibnamefont {Peaudecerf}}, \bibinfo {author}
  {\bibfnamefont {G.~D.}\ \bibnamefont {Bruce}}, \ and\ \bibinfo {author}
  {\bibfnamefont {S.}~\bibnamefont {Kuhr}},\ }\href {\doibase
  10.1038/nphys3403} {\bibfield  {journal} {\bibinfo  {journal} {Nat. Phys.}\
  }\textbf {\bibinfo {volume} {11}},\ \bibinfo {pages} {738} (\bibinfo {year}
  {2015})}\BibitemShut {NoStop}%
\bibitem [{\citenamefont {Edge}\ \emph {et~al.}(2015)\citenamefont {Edge},
  \citenamefont {Anderson}, \citenamefont {Jervis}, \citenamefont {McKay},
  \citenamefont {Day}, \citenamefont {Trotzky},\ and\ \citenamefont
  {Thywissen}}]{Edge2015}%
  \BibitemOpen
  \bibfield  {author} {\bibinfo {author} {\bibfnamefont {G.~J.~A.}\
  \bibnamefont {Edge}}, \bibinfo {author} {\bibfnamefont {R.}~\bibnamefont
  {Anderson}}, \bibinfo {author} {\bibfnamefont {D.}~\bibnamefont {Jervis}},
  \bibinfo {author} {\bibfnamefont {D.~C.}\ \bibnamefont {McKay}}, \bibinfo
  {author} {\bibfnamefont {R.}~\bibnamefont {Day}}, \bibinfo {author}
  {\bibfnamefont {S.}~\bibnamefont {Trotzky}}, \ and\ \bibinfo {author}
  {\bibfnamefont {J.~H.}\ \bibnamefont {Thywissen}},\ }\href {\doibase
  10.1103/PhysRevA.92.063406} {\bibfield  {journal} {\bibinfo  {journal} {Phys.
  Rev. A}\ }\textbf {\bibinfo {volume} {92}},\ \bibinfo {pages} {063406}
  (\bibinfo {year} {2015})}\BibitemShut {NoStop}%
\bibitem [{\citenamefont {Omran}\ \emph {et~al.}(2015)\citenamefont {Omran},
  \citenamefont {Boll}, \citenamefont {Hilker}, \citenamefont {Kleinlein},
  \citenamefont {Salomon}, \citenamefont {Bloch},\ and\ \citenamefont
  {Gross}}]{Omran2015}%
  \BibitemOpen
  \bibfield  {author} {\bibinfo {author} {\bibfnamefont {A.}~\bibnamefont
  {Omran}}, \bibinfo {author} {\bibfnamefont {M.}~\bibnamefont {Boll}},
  \bibinfo {author} {\bibfnamefont {T.~A.}\ \bibnamefont {Hilker}}, \bibinfo
  {author} {\bibfnamefont {K.}~\bibnamefont {Kleinlein}}, \bibinfo {author}
  {\bibfnamefont {G.}~\bibnamefont {Salomon}}, \bibinfo {author} {\bibfnamefont
  {I.}~\bibnamefont {Bloch}}, \ and\ \bibinfo {author} {\bibfnamefont
  {C.}~\bibnamefont {Gross}},\ }\href {\doibase 10.1103/PhysRevLett.115.263001}
  {\bibfield  {journal} {\bibinfo  {journal} {Phys. Rev. Lett.}\ }\textbf
  {\bibinfo {volume} {115}},\ \bibinfo {pages} {263001} (\bibinfo {year}
  {2015})}\BibitemShut {NoStop}%
\bibitem [{\citenamefont {Parsons}\ \emph {et~al.}(2015)\citenamefont
  {Parsons}, \citenamefont {Huber}, \citenamefont {Mazurenko}, \citenamefont
  {Chiu}, \citenamefont {Setiawan}, \citenamefont {Wooley-Brown}, \citenamefont
  {Blatt},\ and\ \citenamefont {Greiner}}]{Parsons2015}%
  \BibitemOpen
  \bibfield  {author} {\bibinfo {author} {\bibfnamefont {M.~F.}\ \bibnamefont
  {Parsons}}, \bibinfo {author} {\bibfnamefont {F.}~\bibnamefont {Huber}},
  \bibinfo {author} {\bibfnamefont {A.}~\bibnamefont {Mazurenko}}, \bibinfo
  {author} {\bibfnamefont {C.~S.}\ \bibnamefont {Chiu}}, \bibinfo {author}
  {\bibfnamefont {W.}~\bibnamefont {Setiawan}}, \bibinfo {author}
  {\bibfnamefont {K.}~\bibnamefont {Wooley-Brown}}, \bibinfo {author}
  {\bibfnamefont {S.}~\bibnamefont {Blatt}}, \ and\ \bibinfo {author}
  {\bibfnamefont {M.}~\bibnamefont {Greiner}},\ }\href {\doibase
  10.1103/PhysRevLett.114.213002} {\bibfield  {journal} {\bibinfo  {journal}
  {Phys. Rev. Lett.}\ }\textbf {\bibinfo {volume} {114}},\ \bibinfo {pages}
  {213002} (\bibinfo {year} {2015})}\BibitemShut {NoStop}%
\bibitem [{\citenamefont {Cheuk}\ \emph {et~al.}(2015)\citenamefont {Cheuk},
  \citenamefont {Nichols}, \citenamefont {Okan}, \citenamefont {Gersdorf},
  \citenamefont {Ramasesh}, \citenamefont {Bakr}, \citenamefont {Lompe},\ and\
  \citenamefont {Zwierlein}}]{Cheuk2015}%
  \BibitemOpen
  \bibfield  {author} {\bibinfo {author} {\bibfnamefont {L.~W.}\ \bibnamefont
  {Cheuk}}, \bibinfo {author} {\bibfnamefont {M.~A.}\ \bibnamefont {Nichols}},
  \bibinfo {author} {\bibfnamefont {M.}~\bibnamefont {Okan}}, \bibinfo {author}
  {\bibfnamefont {T.}~\bibnamefont {Gersdorf}}, \bibinfo {author}
  {\bibfnamefont {V.~V.}\ \bibnamefont {Ramasesh}}, \bibinfo {author}
  {\bibfnamefont {W.~S.}\ \bibnamefont {Bakr}}, \bibinfo {author}
  {\bibfnamefont {T.}~\bibnamefont {Lompe}}, \ and\ \bibinfo {author}
  {\bibfnamefont {M.~W.}\ \bibnamefont {Zwierlein}},\ }\href {\doibase
  10.1103/PhysRevLett.114.193001} {\bibfield  {journal} {\bibinfo  {journal}
  {Phys. Rev. Lett.}\ }\textbf {\bibinfo {volume} {114}},\ \bibinfo {pages}
  {193001} (\bibinfo {year} {2015})}\BibitemShut {NoStop}%
\bibitem [{\citenamefont {Yamamoto}\ \emph {et~al.}(2016)\citenamefont
  {Yamamoto}, \citenamefont {Kobayashi}, \citenamefont {Kuno}, \citenamefont
  {K.},\ and\ \citenamefont {Takahashi}}]{Yamamoto2016}%
  \BibitemOpen
  \bibfield  {author} {\bibinfo {author} {\bibfnamefont {R.}~\bibnamefont
  {Yamamoto}}, \bibinfo {author} {\bibfnamefont {J.}~\bibnamefont {Kobayashi}},
  \bibinfo {author} {\bibfnamefont {T.}~\bibnamefont {Kuno}}, \bibinfo {author}
  {\bibfnamefont {K.}~\bibnamefont {K.}}, \ and\ \bibinfo {author}
  {\bibfnamefont {Y.}~\bibnamefont {Takahashi}},\ }\href@noop {} {\bibfield
  {journal} {\bibinfo  {journal} {New J. Phys.}\ }\textbf {\bibinfo {volume}
  {18}},\ \bibinfo {pages} {023016} (\bibinfo {year} {2016})}\BibitemShut
  {NoStop}%
\bibitem [{\citenamefont {Fulde}\ and\ \citenamefont
  {Ferrell}(1964)}]{Fulde1964}%
  \BibitemOpen
  \bibfield  {author} {\bibinfo {author} {\bibfnamefont {P.}~\bibnamefont
  {Fulde}}\ and\ \bibinfo {author} {\bibfnamefont {R.~A.}\ \bibnamefont
  {Ferrell}},\ }\href {\doibase 10.1103/PhysRev.135.A550} {\bibfield  {journal}
  {\bibinfo  {journal} {Phys. Rev.}\ }\textbf {\bibinfo {volume} {135}},\
  \bibinfo {pages} {A550} (\bibinfo {year} {1964})}\BibitemShut {NoStop}%
\bibitem [{\citenamefont {Z\"urn}\ \emph {et~al.}(2013)\citenamefont {Z\"urn},
  \citenamefont {Lompe}, \citenamefont {Wenz}, \citenamefont {Jochim},
  \citenamefont {Julienne},\ and\ \citenamefont {Hutson}}]{Zurn2013}%
  \BibitemOpen
  \bibfield  {author} {\bibinfo {author} {\bibfnamefont {G.}~\bibnamefont
  {Z\"urn}}, \bibinfo {author} {\bibfnamefont {T.}~\bibnamefont {Lompe}},
  \bibinfo {author} {\bibfnamefont {A.~N.}\ \bibnamefont {Wenz}}, \bibinfo
  {author} {\bibfnamefont {S.}~\bibnamefont {Jochim}}, \bibinfo {author}
  {\bibfnamefont {P.~S.}\ \bibnamefont {Julienne}}, \ and\ \bibinfo {author}
  {\bibfnamefont {J.~M.}\ \bibnamefont {Hutson}},\ }\href {\doibase
  10.1103/PhysRevLett.110.135301} {\bibfield  {journal} {\bibinfo  {journal}
  {Phys. Rev. Lett.}\ }\textbf {\bibinfo {volume} {110}},\ \bibinfo {pages}
  {135301} (\bibinfo {year} {2013})}\BibitemShut {NoStop}%
\bibitem [{\citenamefont {Mitra}\ \emph {et~al.}(2016)\citenamefont {Mitra},
  \citenamefont {Brown}, \citenamefont {Schau\ss{}}, \citenamefont {Kondov},\
  and\ \citenamefont {Bakr}}]{Mitra2016}%
  \BibitemOpen
  \bibfield  {author} {\bibinfo {author} {\bibfnamefont {D.}~\bibnamefont
  {Mitra}}, \bibinfo {author} {\bibfnamefont {P.~T.}\ \bibnamefont {Brown}},
  \bibinfo {author} {\bibfnamefont {P.}~\bibnamefont {Schau\ss{}}}, \bibinfo
  {author} {\bibfnamefont {S.~S.}\ \bibnamefont {Kondov}}, \ and\ \bibinfo
  {author} {\bibfnamefont {W.~S.}\ \bibnamefont {Bakr}},\ }\href {\doibase
  10.1103/PhysRevLett.117.093601} {\bibfield  {journal} {\bibinfo  {journal}
  {Phys. Rev. Lett.}\ }\textbf {\bibinfo {volume} {117}},\ \bibinfo {pages}
  {093601} (\bibinfo {year} {2016})}\BibitemShut {NoStop}%
\bibitem [{\citenamefont {Ho}\ \emph {et~al.}(2009)\citenamefont {Ho},
  \citenamefont {Cazalilla},\ and\ \citenamefont {Giamarchi}}]{Ho2009}%
  \BibitemOpen
  \bibfield  {author} {\bibinfo {author} {\bibfnamefont {A.~F.}\ \bibnamefont
  {Ho}}, \bibinfo {author} {\bibfnamefont {M.~A.}\ \bibnamefont {Cazalilla}}, \
  and\ \bibinfo {author} {\bibfnamefont {T.}~\bibnamefont {Giamarchi}},\ }\href
  {\doibase 10.1103/PhysRevA.79.033620} {\bibfield  {journal} {\bibinfo
  {journal} {Phys. Rev. A}\ }\textbf {\bibinfo {volume} {79}},\ \bibinfo
  {pages} {033620} (\bibinfo {year} {2009})}\BibitemShut {NoStop}%
\bibitem [{\citenamefont {Varney}\ \emph {et~al.}(2009)\citenamefont {Varney},
  \citenamefont {Lee}, \citenamefont {Bai}, \citenamefont {Chiesa},
  \citenamefont {Jarrell},\ and\ \citenamefont {Scalettar}}]{Varney2009}%
  \BibitemOpen
  \bibfield  {author} {\bibinfo {author} {\bibfnamefont {C.~N.}\ \bibnamefont
  {Varney}}, \bibinfo {author} {\bibfnamefont {C.-R.}\ \bibnamefont {Lee}},
  \bibinfo {author} {\bibfnamefont {Z.~J.}\ \bibnamefont {Bai}}, \bibinfo
  {author} {\bibfnamefont {S.}~\bibnamefont {Chiesa}}, \bibinfo {author}
  {\bibfnamefont {M.}~\bibnamefont {Jarrell}}, \ and\ \bibinfo {author}
  {\bibfnamefont {R.~T.}\ \bibnamefont {Scalettar}},\ }\href {\doibase
  10.1103/PhysRevB.80.075116} {\bibfield  {journal} {\bibinfo  {journal} {Phys.
  Rev. B}\ }\textbf {\bibinfo {volume} {80}},\ \bibinfo {pages} {075116}
  (\bibinfo {year} {2009})}\BibitemShut {NoStop}%
\bibitem [{\citenamefont {Demler}\ \emph {et~al.}(1996)\citenamefont {Demler},
  \citenamefont {Zhang}, \citenamefont {Bulut},\ and\ \citenamefont
  {Scalapino}}]{Demler1996}%
  \BibitemOpen
  \bibfield  {author} {\bibinfo {author} {\bibfnamefont {E.}~\bibnamefont
  {Demler}}, \bibinfo {author} {\bibfnamefont {S.-C.}\ \bibnamefont {Zhang}},
  \bibinfo {author} {\bibfnamefont {N.}~\bibnamefont {Bulut}}, \ and\ \bibinfo
  {author} {\bibfnamefont {D.~J.}\ \bibnamefont {Scalapino}},\ }\href {\doibase
  10.1142/S0217979296000982} {\bibfield  {journal} {\bibinfo  {journal} {Int.
  J. Mod. Phys. B}\ }\textbf {\bibinfo {volume} {10}},\ \bibinfo {pages} {2137}
  (\bibinfo {year} {1996})}\BibitemShut {NoStop}%
\bibitem [{\citenamefont {Moreo}\ and\ \citenamefont
  {Scalapino}(2007)}]{Moreo2007}%
  \BibitemOpen
  \bibfield  {author} {\bibinfo {author} {\bibfnamefont {A.}~\bibnamefont
  {Moreo}}\ and\ \bibinfo {author} {\bibfnamefont {D.~J.}\ \bibnamefont
  {Scalapino}},\ }\href {\doibase 10.1103/PhysRevLett.98.216402} {\bibfield
  {journal} {\bibinfo  {journal} {Phys. Rev. Lett.}\ }\textbf {\bibinfo
  {volume} {98}},\ \bibinfo {pages} {216402} (\bibinfo {year}
  {2007})}\BibitemShut {NoStop}%
\bibitem [{\citenamefont {Loh}\ and\ \citenamefont {Trivedi}(2010)}]{Loh2010}%
  \BibitemOpen
  \bibfield  {author} {\bibinfo {author} {\bibfnamefont {Y.~L.}\ \bibnamefont
  {Loh}}\ and\ \bibinfo {author} {\bibfnamefont {N.}~\bibnamefont {Trivedi}},\
  }\href {\doibase 10.1103/PhysRevLett.104.165302} {\bibfield  {journal}
  {\bibinfo  {journal} {Phys. Rev. Lett.}\ }\textbf {\bibinfo {volume} {104}},\
  \bibinfo {pages} {165302} (\bibinfo {year} {2010})}\BibitemShut {NoStop}%
\bibitem [{\citenamefont {Kim}\ and\ \citenamefont
  {T\"orm\"a}(2012)}]{Kim2012}%
  \BibitemOpen
  \bibfield  {author} {\bibinfo {author} {\bibfnamefont {D.-H.}\ \bibnamefont
  {Kim}}\ and\ \bibinfo {author} {\bibfnamefont {P.}~\bibnamefont
  {T\"orm\"a}},\ }\href {\doibase 10.1103/PhysRevB.85.180508} {\bibfield
  {journal} {\bibinfo  {journal} {Phys. Rev. B}\ }\textbf {\bibinfo {volume}
  {85}},\ \bibinfo {pages} {180508} (\bibinfo {year} {2012})}\BibitemShut
  {NoStop}%
\bibitem [{\citenamefont {Gukelberger}\ \emph {et~al.}(2016)\citenamefont
  {Gukelberger}, \citenamefont {Lienert}, \citenamefont {Kozik}, \citenamefont
  {Pollet},\ and\ \citenamefont {Troyer}}]{Gukelberger2016}%
  \BibitemOpen
  \bibfield  {author} {\bibinfo {author} {\bibfnamefont {J.}~\bibnamefont
  {Gukelberger}}, \bibinfo {author} {\bibfnamefont {S.}~\bibnamefont
  {Lienert}}, \bibinfo {author} {\bibfnamefont {E.}~\bibnamefont {Kozik}},
  \bibinfo {author} {\bibfnamefont {L.}~\bibnamefont {Pollet}}, \ and\ \bibinfo
  {author} {\bibfnamefont {M.}~\bibnamefont {Troyer}},\ }\href {\doibase
  10.1103/PhysRevB.94.075157} {\bibfield  {journal} {\bibinfo  {journal} {Phys.
  Rev. B}\ }\textbf {\bibinfo {volume} {94}},\ \bibinfo {pages} {075157}
  (\bibinfo {year} {2016})}\BibitemShut {NoStop}%
\bibitem [{\citenamefont {Mitra}\ \emph {et~al.}(2016)\citenamefont {Mitra},
  \citenamefont {Brown}, \citenamefont {Schau\ss{}}, \citenamefont {Kondov},\
  and\ \citenamefont {Bakr}}]{Mitra2016}%
  \BibitemOpen
  \bibfield  {author} {\bibinfo {author} {\bibfnamefont {D.}~\bibnamefont
  {Mitra}}, \bibinfo {author} {\bibfnamefont {P.~T.}\ \bibnamefont {Brown}},
  \bibinfo {author} {\bibfnamefont {P.}~\bibnamefont {Schau\ss{}}}, \bibinfo
  {author} {\bibfnamefont {S.~S.}\ \bibnamefont {Kondov}}, \ and\ \bibinfo
  {author} {\bibfnamefont {W.~S.}\ \bibnamefont {Bakr}},\ }\href {\doibase
  10.1103/PhysRevLett.117.093601} {\bibfield  {journal} {\bibinfo  {journal}
  {Phys. Rev. Lett.}\ }\textbf {\bibinfo {volume} {117}},\ \bibinfo {pages}
  {093601} (\bibinfo {year} {2016})}\BibitemShut {NoStop}%
\bibitem [{\citenamefont {Idziaszek}\ and\ \citenamefont
  {Calarco}(2005)}]{Idziaszek2005}%
  \BibitemOpen
  \bibfield  {author} {\bibinfo {author} {\bibfnamefont {Z.}~\bibnamefont
  {Idziaszek}}\ and\ \bibinfo {author} {\bibfnamefont {T.}~\bibnamefont
  {Calarco}},\ }\href {\doibase 10.1103/PhysRevA.71.050701} {\bibfield
  {journal} {\bibinfo  {journal} {Phys. Rev. A}\ }\textbf {\bibinfo {volume}
  {71}},\ \bibinfo {pages} {050701} (\bibinfo {year} {2005})}\BibitemShut
  {NoStop}%
\bibitem [{\citenamefont {Zhang}(1990)}]{Zhang1990}%
  \BibitemOpen
  \bibfield  {author} {\bibinfo {author} {\bibfnamefont {S.}~\bibnamefont
  {Zhang}},\ }\href {\doibase 10.1103/PhysRevLett.65.120} {\bibfield  {journal}
  {\bibinfo  {journal} {Phys. Rev. Lett.}\ }\textbf {\bibinfo {volume} {65}},\
  \bibinfo {pages} {120} (\bibinfo {year} {1990})}\BibitemShut {NoStop}%
\bibitem [{\citenamefont {Ho}\ \emph {et~al.}(2009)\citenamefont {Ho},
  \citenamefont {Cazalilla},\ and\ \citenamefont {Giamarchi}}]{Ho2009}%
  \BibitemOpen
  \bibfield  {author} {\bibinfo {author} {\bibfnamefont {A.~F.}\ \bibnamefont
  {Ho}}, \bibinfo {author} {\bibfnamefont {M.~A.}\ \bibnamefont {Cazalilla}}, \
  and\ \bibinfo {author} {\bibfnamefont {T.}~\bibnamefont {Giamarchi}},\ }\href
  {\doibase 10.1103/PhysRevA.79.033620} {\bibfield  {journal} {\bibinfo
  {journal} {Phys. Rev. A}\ }\textbf {\bibinfo {volume} {79}},\ \bibinfo
  {pages} {033620} (\bibinfo {year} {2009})}\BibitemShut {NoStop}%
\bibitem [{\citenamefont {J\"urgensen}\ \emph {et~al.}(2014)\citenamefont
  {J\"urgensen}, \citenamefont {Meinert}, \citenamefont {Mark}, \citenamefont
  {N\"agerl},\ and\ \citenamefont {L\"uhmann}}]{Jurgensen2014}%
  \BibitemOpen
  \bibfield  {author} {\bibinfo {author} {\bibfnamefont {O.}~\bibnamefont
  {J\"urgensen}}, \bibinfo {author} {\bibfnamefont {F.}~\bibnamefont
  {Meinert}}, \bibinfo {author} {\bibfnamefont {M.~J.}\ \bibnamefont {Mark}},
  \bibinfo {author} {\bibfnamefont {H.-C.}\ \bibnamefont {N\"agerl}}, \ and\
  \bibinfo {author} {\bibfnamefont {D.-S.}\ \bibnamefont {L\"uhmann}},\ }\href
  {\doibase 10.1103/PhysRevLett.113.193003} {\bibfield  {journal} {\bibinfo
  {journal} {Phys. Rev. Lett.}\ }\textbf {\bibinfo {volume} {113}},\ \bibinfo
  {pages} {193003} (\bibinfo {year} {2014})}\BibitemShut {NoStop}%
\bibitem [{\citenamefont {Varney}\ \emph {et~al.}(2009)\citenamefont {Varney},
  \citenamefont {Lee}, \citenamefont {Bai}, \citenamefont {Chiesa},
  \citenamefont {Jarrell},\ and\ \citenamefont {Scalettar}}]{Varney2009}%
  \BibitemOpen
  \bibfield  {author} {\bibinfo {author} {\bibfnamefont {C.~N.}\ \bibnamefont
  {Varney}}, \bibinfo {author} {\bibfnamefont {C.-R.}\ \bibnamefont {Lee}},
  \bibinfo {author} {\bibfnamefont {Z.~J.}\ \bibnamefont {Bai}}, \bibinfo
  {author} {\bibfnamefont {S.}~\bibnamefont {Chiesa}}, \bibinfo {author}
  {\bibfnamefont {M.}~\bibnamefont {Jarrell}}, \ and\ \bibinfo {author}
  {\bibfnamefont {R.~T.}\ \bibnamefont {Scalettar}},\ }\href {\doibase
  10.1103/PhysRevB.80.075116} {\bibfield  {journal} {\bibinfo  {journal} {Phys.
  Rev. B}\ }\textbf {\bibinfo {volume} {80}},\ \bibinfo {pages} {075116}
  (\bibinfo {year} {2009})}\BibitemShut {NoStop}%
\bibitem [{\citenamefont {Assmann}\ \emph {et~al.}(2012)\citenamefont
  {Assmann}, \citenamefont {Chiesa}, \citenamefont {Batrouni}, \citenamefont
  {Evertz},\ and\ \citenamefont {Scalettar}}]{Assmann2012}%
  \BibitemOpen
  \bibfield  {author} {\bibinfo {author} {\bibfnamefont {E.}~\bibnamefont
  {Assmann}}, \bibinfo {author} {\bibfnamefont {S.}~\bibnamefont {Chiesa}},
  \bibinfo {author} {\bibfnamefont {G.~G.}\ \bibnamefont {Batrouni}}, \bibinfo
  {author} {\bibfnamefont {H.~G.}\ \bibnamefont {Evertz}}, \ and\ \bibinfo
  {author} {\bibfnamefont {R.~T.}\ \bibnamefont {Scalettar}},\ }\href {\doibase
  10.1103/PhysRevB.85.014509} {\bibfield  {journal} {\bibinfo  {journal} {Phys.
  Rev. B}\ }\textbf {\bibinfo {volume} {85}},\ \bibinfo {pages} {014509}
  (\bibinfo {year} {2012})}\BibitemShut {NoStop}%
\end{thebibliography}
\end{document}